\documentstyle[preprint,aps]{revtex}
\tighten
\draft
\widetext
\input epsf
\preprint{HUTP-98/A031, NUB 3178}
\bigskip
\bigskip
\begin{document}
\title{On Four Dimensional ${\cal N}=1$ Type I Compactifications}
\medskip
\author{Zurab Kakushadze\footnote{E-mail: 
zurab@string.harvard.edu}}
\bigskip
\address{Lyman Laboratory of Physics, Harvard University, Cambridge, 
MA 02138\\
and\\
Department of Physics, Northeastern University, Boston, MA 02115}
\date{June 1, 1998}
\bigskip
\medskip
\maketitle

\begin{abstract}
We consider four dimensional ${\cal N}=1$ supersymmetric Type I compactifications on
toroidal orbifolds $T^6/\Gamma$. In particular, we focus on the Type I vacua which are perturbative from the orientifold viewpoint, that is, on the compactifications with well defined
world-sheet expansion. The number of such models is rather constrained. This allows us
to study all such vacua. This, in particular, involves considering compactifications with non-trivial NS-NS antisymmetric tensor backgrounds. We derive massless spectra for these models, and also compute superpotentials. We review the reasons responsible for such a limited
number of perturbative Type I compactifications on toroidal orbifolds (which include Abelian as well as non-Abelian cases). As an aside, we generalize the recent work on large $N$ gauge theories from orientifolds to include a non-Abelian orbifold. This also provides an important independent check for perturbative consistency of the corresponding Type I compactification.
\end{abstract}

\section{Introduction}

{}In recent years substantial progress has been made in understanding non-perturbative
string dynamics. In particular, in ten dimensions there are five consistent superstring theories.
The first four, Type IIA, Type IIB, $E_8\otimes E_8$ heterotic and ${\mbox{Spin}}(32)/{\bf Z}_2$
heterotic, are theories of oriented closed strings. Type I superstring is a theory of both unoriented
closed plus open strings. Perturbatively, these five theories are distinct. Nonetheless, these
theories exhibit a web of (conjectured) dualities which all seem to point to an underlying unified
description. Most of these dualities are intrinsically non-perturbative and often allow to
map non-perturbative phenomena in one theory to perturbative phenomena in another theory.

{}The four
oriented closed string theories and their compactifications are relatively well understood as far as perturbative formulation is concerned. Conformal field theory and modular invariance serve as guiding principles for 
perturbative model building in closed string theories. Type I, however, lacks modular invariance.
Moreover, conformal field theories on world-sheets with boundaries (invariably present
in open string theories) are still poorly understood. These have been some of the main reasons
for lack of as deep understanding of perturbative Type I compactifications as in closed string theories.

{}In the past years various unoriented closed plus open string vacua have been
constructed using orientifold techniques. Type IIB orientifolds are 
generalized orbifolds 
that involve world-sheet parity reversal along with geometric symmetries 
of the theory \cite{group}. Orientifolding procedure results in an unoriented closed string theory.
Consistency then generically requires introducing open strings that can 
be viewed as starting and ending on D-branes \cite{Db}. 
In particular, Type I compactifications on toroidal orbifolds can be viewed as Type IIB
orientifolds with a certain choice of the orientifold projection. 
Global Chan-Paton charges
associated with D-branes manifest themselves as a gauge symmetry in 
space-time. The orientifold techniques have been successfully
applied to the construction of six dimensional ${\cal N}=1$ space-time supersymmetric 
orientifolds of Type IIB compactified on orbifold limits of K3 (that is, toroidal orbifolds
$T^4/{\bf Z}_N$, $N=2,3,4,6$) \cite{PS,GJ,NS}. 
These orientifold models generically contain more than one tensor multiplet and/or enhanced gauge symmetries from D5-branes in their massless spectra,
and, therefore, describe six dimensional vacua which are non-perturbative from the 
heterotic viewpoint.

{}The orientifold construction has subsequently been generalized to four dimensional ${\cal N}=1$ space-time supersymmetric compactifications \cite{BL,Sagnotti,ZK,KS1,KS2,zk}. Several
such orientifolds have been constructed. Some of these models, namely, the ${\bf Z}_3$ \cite{Sagnotti}, ${\bf Z}_7$ \cite{KS1} and ${\bf Z}_3\otimes {\bf Z}_3$ \cite{KS2} orbifold
models have perturbative heterotic duals \cite{ZK,KS1,KS2}. Others, such as the ${\bf Z}_2\otimes {\bf Z}_2$ \cite{BL}, ${\bf Z}_6$ \cite{KS2} and 
${\bf Z}_2\otimes {\bf Z}_2\otimes {\bf Z}_3$ \cite{zk}
orbifold models are non-perturbative
from the heterotic viewpoint \cite{ZK,KS1,KS2} as they contain D5-branes. In particular, the  
${\bf Z}_6$ orbifold model of \cite{KS2} and the 
${\bf Z}_2\otimes {\bf Z}_2\otimes {\bf Z}_3$ model of \cite{zk}
are the first known examples of consistent chiral ${\cal N}=1$ string vacua in four dimensions that are non-perturbative from the orientifold viewpoint.

{}Despite the above developments at some point it became clear that our understanding of 
orientifolds was incomplete. In particular, in some of the models discussed in \cite{Zwart,AFIV}
the tadpole cancellation conditions (derived using the perturbative orientifold
approach, namely, via a straightforward generalization of the six dimensional 
tadpole cancellation conditions to four dimensions) allowed for no solutions.
This appeared to be a distress signal indicating that a
better understanding of the orientifold construction was called for.

{}Recently, some progress has been made in this direction \cite{KST}. In particular,
in \cite{KST} conditions necessary for world-sheet
consistency of six and four dimensional ${\cal N}=1$ supersymmetric Type IIB 
orientifolds were studied. It was argued that in most cases orientifolds contain sectors
which are non-perturbative ({\em i.e.}, these sectors have no world-sheet description). These
sectors can be thought of as arising from D-branes wrapping various collapsed 2-cycles
in the orbifold. In particular, such non-perturbative states are present
in the ``anomalous'' models of Ref \cite{Zwart} (as well as in other examples of this type
recently discussed in Ref \cite{AFIV}). This resolves the corresponding ``puzzles''.
Certain world-sheet consistency conditions in four dimensional cases
(which are automatically satisfied in the six dimensions)
were pointed out in \cite{KST}
which indicate that the only four dimensional (Abelian) 
orientifolds that have perturbative description
are those of Type IIB compactified on the ${\bf Z}_2\otimes {\bf Z}_2$ \cite{BL}, ${\bf Z}_3$
\cite{Sagnotti}, ${\bf Z}_7$ \cite{KS1}, ${\bf Z}_3\otimes {\bf Z}_3$ and ${\bf Z}_6$
\cite{KS2}, and ${\bf Z}_2\otimes{\bf Z}_2 \otimes {\bf Z}_3$ \cite{zk}. (In this paper we will add one more orbifold group to this list. This is the only non-Abelian orbifold group for which the corresponding orientifold has a world-sheet description.) In particular, 
none of the other models considered in \cite{Zwart,AFIV} have perturbative orientifold
description, and even in the models with all tadpoles cancelled the massless spectra given
in \cite{Zwart,AFIV} miss certain non-perturbative states. These conclusions of \cite{KST} are supported by various checks performed in \cite{KST} and \cite{zura1}. In particular, using Type I-heterotic duality \cite{PW} in the spirit of \cite{ZK,KS1,KS2} as well as F-theory \cite{vafa} picture, it
was possible to determine the world-sheet consistency conditions, that is, the conditions under which the non-perturbative states from wrapped D-branes are either absent or become massive and decouple from the massless spectrum. This gave rise to the list of orbifolds mentioned above. On the other hand, \cite{zura1} utilized the fact that the world-sheet consistency conditions are local statements about orientifold planes and D-branes near orbifold singularities (as far
as geometry is concerned) to perform independent checks of the conclusions in \cite{KST}. These {\em a priori} independent checks appear to be rather convincing as they are in one-to-one correspondence with each other.

{}Although the number of perturbative Type IIB orientifolds in four dimensions appears to be
rather constrained, it is important to understand all the cases at hand. The cases of most interest
are those with D5-branes as they correspond to non-perturbative heterotic vacua. 
In particular, ${\cal N}=1$ orientifolds have mostly been studied in the context of zero NS-NS
$B$-field background. In this paper we study all perturbative ${\cal N}=1$ orientifolds with and without the
$B$-field backgrounds. We systematically classify these models which includes the derivation
of their complete massless spectra. We also compute the renormalizable couplings in the superpotentials of these models. This paper therefore completes the program of constructing and understanding such compactifications.

{}In Fig.1 we have drawn a schematic picture of the space of four dimensional ${\cal N}=1$ Type I and
heterotic vacua. The region ${\cal A}\cup{\cal B}$ corresponds to perturbative Type I vacua.
The region ${\cal A}\cup{\cal C}$ corresponds to perturbative heterotic vacua. The vacua
in the region ${\cal A}$ are perturbative from both the Type I and heterotic viewpoints. The region ${\cal D}$ contains both non-perturbative Type I and heterotic vacua. In this paper we are
concentrating on the region ${\cal A}\cup{\cal B}$. The region ${\cal A}\cup{\cal C}$ is relatively well understood. The ultimate challenge is to understand compactifications which are non-perturbative from both the Type I and heterotic viewpoints, that is, the region ${\cal D}$. It should contain the bulk of interesting Type I/heterotic vacua. Understanding such compactifications would certainly shed some light on the underlying unifying structure of string theory.

\section{Preliminaries}

{}In this section we briefly review some useful facts about orientifolds. In particular, we 
discuss the effect of non-zero NS-NS $B$-field. 

{}Consider Type IIB string theory compactified on ${\cal M}=T^6/\Gamma$ where 
the orbifold group
$\Gamma=\{g_a | a=1,\dots,|\Gamma|\}$ ($g_1=1$) is a finite discrete subgroup
of $SU(3)$ that acts crystallographically on $T^6$. 
Note that ${\cal M}$ is an orbifold Calabi-Yau three-fold with $SU(3)$ holonomy.
The resulting four dimensional theory has ${\cal N}=2$ supersymmetry.
Consider the $\Omega$ orientifold of this theory, where $\Omega$ is the world-sheet parity
reversal. The orientifold group is then ${\cal O}=\{g_a, \Omega g_a | a=1,\dots,|\Gamma|\}$.
The orientifold theory has ${\cal N}=1$ supersymmetry in four dimensions. The unoriented closed string sector contains $h^{1,1}+h^{2,1}$ chiral supermultiplets\footnote{Note that
these chiral supermultiplets are neutral under the Chan-Paton gauge group coming from the open string sector. More precisely, they generically have non-trivial transformation property under anomalous $U(1)$'s (if any) but are neutral with respect to other subgroups of the Chan-Paton gauge group - see below for more details.}, where $(h^{1,1},h^{2,1})$
are the corresponding Hodge numbers of the Calabi-Yau three-fold ${\cal M}$.   

{}Let us first discuss the cases with zero NS-NS $B$-field ($B_{ij}=0$, $i,j=1,\dots,6$). 
Note that we have an orientifold 9-plane corresponding to the element $\Omega$ of the orientifold group. To cancel the corresponding R-R charge we must introduce 32 D9-branes.
(The number 32 of D9-branes is required by the corresponding {\em untwisted} tadpole
cancellation conditions.) If $\Gamma$ has a ${\bf Z}_2$ subgroup, then we also have an
orientifold 5-plane. This corresponds to the element $\Omega R$ of the orientifold group, where
$R$ is the generator of this ${\bf Z}_2$ subgroup. There is a different set of D5-branes corresponding to each ${\bf Z}_2$ subgroup of $\Gamma$. Each set consists of 32 D5-branes.
(The number 32 of D5-branes is required by the corresponding tadpole
cancellation conditions. This also follows from T-duality between D9- and D5-branes.)

{}We need to specify the action of the orbifold group $\Gamma$ on the Chan-Paton factors corresponding to the D9- and D5-branes. This action is given by the corresponding Chan-Paton
matrices which we collectively refer to
as $\gamma^\mu_a$, where the superscript $\mu$ refers to the corresponding
D9- or D5-branes. Note that ${\mbox{Tr}}(\gamma^\mu_1)=n^\mu$ where 
$n^\mu$ is the number of D-branes labelled by $\mu$.

{}The world-sheet consistency of an orientifold theory requires that all massless tadpoles
cancel. These tadpoles arise at one-loop level from the following three sources:
the Klein bottle, annulus, and M{\"o}bius strip amplitudes. The factorization property of string theory implies that the tadpole cancellation conditions read (see, {\em e.g.}, \cite{zura}
for a more detailed discussion):
\begin{equation}\label{BC}
 B_a+\sum_\mu C^\mu_a {\mbox{Tr}}(\gamma^\mu_a)=0~.
\end{equation}
Here $B_a$ and $C^\mu_a$ are (model dependent) numerical coefficients of order 1. 
In the following we will explicitly give the solutions of these tadpole cancellation conditions
for each model. (A detailed discussion of the tadpole cancellation conditions used in this 
paper can be found in \cite{KS1,KS2,zura,zura1}.)

{}Naively, it might seem that the $\Omega$ orientifolds of Type IIB on $T^6/\Gamma$ should have a world-sheet description for any choice of $\Gamma\subset SU(3)$ which acts crystallographically on $T^6$. This is, however, not the case. As discussed at length in \cite{KST}, in most cases orientifolds contain sectors which are non-perturbative from the
orientifold viewpoint, that is, such sectors have no world-sheet description. These sectors
arise from D-branes wrapping various (collapsed) two-cycles in the orbifold. However, using
Type I-heterotic duality \cite{PW} along the lines of \cite{ZK,KS1,KS2} together with F-theory
\cite{vafa} considerations, it is possible to determine for which orbifold groups the corresponding orientifold is perturbative \cite{KST}. The arguments of \cite{KST} are supported by various independent checks performed in \cite{KST} and \cite{zura1}. The number of such orbifold groups turns out rather limited. In particular, there are only six Abelian subgroups of $SU(3)$
for which the corresponding orientifolds have perturbative description. These are the ${\bf Z}_2\otimes {\bf Z}_2$ \cite{BL}, ${\bf Z}_3$ \cite{Sagnotti}, ${\bf Z}_7$ \cite{KS1},
${\bf Z}_3\otimes {\bf Z}_3$ and ${\bf Z}_6$ \cite{KS2}, and ${\bf Z}_2\otimes {\bf Z}_2\otimes {\bf Z}_3$ \cite{zk} cases\footnote{Thus, other orientifold models discussed in \cite{Zwart} and \cite{AFIV} all suffer from additional non-perturbative states.}. There is one other such case which has not been discussed previously: this is a non-Abelian orientifold model considered in section \ref{NA}
of this paper. These seven cases exhaust the list of orbifold groups for which the corresponding
orientifolds can be described perturbatively as world-sheet theories. In the remainder of this paper we will confine our attention to these orbifolds.

{}The above perturbative orientifolds have been mostly discussed in the context of zero NS-NS
$B$-field\footnote{The ${\bf Z}_3$ case with non-zero $B$-field backgrounds has been briefly
discussed in \cite{Sagnotti}. Also, the ${\bf Z}_6$ model of \cite{KS2} with a certain non-zero
$B$-field configuration has been recently discussed in \cite{ST} in a phenomenological context.}. The goal of this paper is to systematically study all perturbative orientifolds with ${\cal N}=1$ supersymmetry in four dimensions\footnote{A systematic classification of six dimensional ${\cal N}=1$ orientifolds with and without the $B$-field has been performed in \cite{NS}.}. Thus, we will also consider the above orientifolds in the presence of non-zero NS-NS $B$-field. 

{}Before we turn to the construction of the corresponding models, let us briefly review the effect of the $B$-field on the orientifold spectrum. Note that the untwisted NS-NS two-form $B_{ij}$ is
odd under the orientifold projection $\Omega$. This implies that the corresponding states are
projected out of the closed string massless spectrum. Nonetheless, certain quantized vacuum expectation values of $B_{ij}$ are allowed. This can be seen as follows. Let $B_{ij}$ be normalized such that it is defined up to a unit shift: $B_{ij}\sim B_{ij}+1$. With this normalization,
the only two values of $B_{ij}$ invariant under $\Omega$ are 0 and $1/2$, hence quantization of
$B_{ij}$. 

{}The effect of quantized $B$-field in toroidal compactifications of Type I string theory has been studied in \cite{Bij} (and also more recently in \cite{bianchi,toroidal}). This discussion has been
recently generalized to six-dimensional orbifold compactifications in \cite{NS}\footnote{Orientifolds of Type IIB on smooth K3 with non-zero $B$-field have been studied in \cite{SS}.}. The results of \cite{NS} can be generalized to four dimensional orbifold compactifications. Here we briefly state the key facts relevant for the subsequent discussions, and refer the reader to \cite{NS} for more details.

{}Thus, let $b$ denote the rank of the $6\times 6$ matrix $B_{ij}$. (Note that due to antisymmetry of $B_{ij}$, its rank $b\in 2{\bf Z}$.) As discussed in \cite{NS}, the numbers of D9- and D5-branes  
(of each type) are $32/2^{b/2}$. (For $b=0$ we have the usual number 32 of each type of branes.) For $b\not=0$ the untwisted tadpole cancellation conditions for the D9-branes {\em a priori} allow the orientifold projection $\Omega$ to be of either $SO$ or $Sp$ type, that is, before taking into account the orbifold projections (and the corresponding twisted tadpole cancellation conditions) the resulting ${\cal N}=4$ toroidal compactification of Type I has the gauge group $SO(32/2^{b/2})$ or $Sp(32/2^{b/2})$. (Here we are using the convention where the rank of
$Sp(2N)$ is $N$.) Note that for $b=0$ only the $SO$ type projection is allowed for D9-branes.
As to the twisted tadpole cancellation conditions, the following modifications occur for $b\not=0$. Suppose $g_a$ is an element of the orbifold group $\Gamma$ such that the fixed point locus of the corresponding twist is of real dimension two. Then, if the components of $B_{ij}$ corresponding to this locus form a non-zero $2\times 2$ matrix (that is, an antisymmetric matrix of rank 2), the coefficient $B_a$ in (\ref{BC}) is {\em reduced} by a factor of 2 (compared
with the case $b=0$). None of the other twisted tadpoles get modified. (However, the type of the  
orientifold projection ({\em i.e.}, whether it is of the $SO$ or $Sp$ type) is fixed by the twisted tadpole cancellation conditions.)

{}Another important point is multiplicity of states in sectors corresponding to open strings stretched between different species of D-branes. Thus, consider a model with D9- plus
D$5_s$-branes. (As we will see in the subsequent sections, there can be only one, 
two or three different types of D5-branes.) The states in the $95_s$ and $5_s 5_s^{\prime}$ ($s\not=s^\prime$) sectors transform in the bi-fundamental irreps of the corresponding gauge groups. These bi-fundamental irreps appear with the multiplicity $k=2^{b/2}$. (Thus, for $b=0$ this multiplicity is one.) This multiplicity of states in these sectors is due to (see below for more details) the $({\bf Z}_2)^{b/2}$ discrete symmetry (which is trivial in the $b=0$ case) present in these models. This discrete symmetry will be important in determining the superpotential in the models with $b\not=0$. 

{}We are now ready to discuss the four dimensional orientifolds with non-zero $B$-field.

\section{Abelian Orbifolds}

{}In this section we will discuss the $\Omega$ orientifolds of Type IIB compactified on $T^6/\Gamma$, where $\Gamma$ is one of the six Abelian orbifolds mentioned in the previous section, namely, ${\bf Z}_2\otimes {\bf Z}_2$, ${\bf Z}_3$, ${\bf Z}_7$,
${\bf Z}_3\otimes {\bf Z}_3$, ${\bf Z}_6$, and ${\bf Z}_2\otimes {\bf Z}_2\otimes {\bf Z}_3$. For completeness we will discuss all the models including those with zero $B$-field. This should make the discussion of the cases with $b\not=0$ easier to follow. For each of these models we will give the {\em renormalizable} (if any) terms in the superpotential. (Higher dimensional terms are not difficult to reproduce.) 

{}In all the cases except ${\bf Z}_7$ we can for simplicity assume that the six-torus $T^6$ factorizes into a product of three two-tori: $T^6=T^2\otimes T^2\otimes T^2$. Let $z_s$, $s=1,2,3$ be the complex coordinates parametrizing these two-tori. Then the orbifold group elements $g_a$ are $3\times 3$ $SU(3)$ matrices acting on $z_s$: $g_a z_s=(g_a)_{ss^\prime}
z_{s^\prime}$. Moreover, for our purposes here it will suffice\footnote{The cases
with non-block-diagonal $B_{ij}$ can be studied similarly and are not difficult to work out. The corresponding open string sectors are the same as in the block-diagonal cases. The closed string sectors are generically different as the Hodge numbers of the corresponding orbifold Calabi-Yau three-fold depend on the precise form of the $B_{ij}$ matrix.} to assume that the matrix $B_{ij}$ is block-diagonal:
$(B_{ij})={\mbox{diag}}(B_{12}\epsilon, B_{34}\epsilon, B_{56}\epsilon)$, where $\epsilon$ 
is an antisymmetric $2\times 2$ matrix with $\epsilon_{12}=-\epsilon_{21}=1$. Here we assume that the internal directions $(1,2)$, $(3,4)$ and $(5,6)$ correspond to the complex coordinates
$z_1$, $z_2$ and $z_3$, respectively. Note that there is no restriction on $B_{12}$, $B_{34}$ and $B_{56}$ other than that they take values 0 or $1/2$.

{}In the ${\bf Z}_7$ case the requirement that the orbifold group act crystallographically on $T^6$ implies that $T^6$ cannot be factorized. Also, the rank of the $B$-field can only be either $b=0$ or $b=6$.

{}We now turn to the detailed description of each model.

\subsection{The ${\bf Z}_3$ Orbifold}

{}Let $g$ be the generator of the 
orbifold group $\Gamma\approx{\bf Z}_3$. The action of $g$ on the
complex coordinates $z_s$ is given by:
\begin{equation}
 gz_s=\omega z_s~,~~~\omega=\exp(2\pi i/3)~.
\end{equation} 
(The corresponding Calabi-Yau three-fold ${\cal M}=T^6/\Gamma$ has the following Hodge numbers: $(h^{1,1},h^{2,1})=(36,0)$. Thus, there are 36 chiral supermultiplets in the closed 
string sector.)
In this case we have D9-branes only.
The Chan-Paton matrices corresponding to the solutions of tadpole cancellation conditions depend on the rank $b$ of the $B_{ij}$ matrix.\\
${\bullet}$ $b=0$. We have 32 D9-branes. The orientifold projection is of the $SO$ type.
The solution to the twisted tadpole cancellation conditions is given by:
\begin{equation}
 \gamma_{g,9}={\mbox{diag}}(\omega {\bf I}_6,\omega^{-1} {\bf I}_6,{\bf I}_4)~.
\end{equation} 
Here (and throughout this paper unless explicitly stated otherwise) we choose not to count the orientifold images of the D-branes.
For this reason, here $\gamma_{g,9}$ is a $16\times 16$ (and not a $32\times 32$) matrix. Also,
${\bf I}_N$ denotes an $N\times N$ identity matrix. The massless spectrum of this model is given in Table I. The gauge group is $U(12)\otimes SO(8)$. (This model was originally constructed in \cite{Sagnotti}.) The superpotential reads (here and in the following we suppress the actual values of the Yukawa couplings and only display the non-vanishing terms, and the summation
over repeated indices is understood):
\begin{equation}
 {\cal W}=\epsilon_{s{s^\prime}s^{\prime\prime}} \Phi_s Q_{s^\prime} Q_{s^{\prime\prime}}~.
\end{equation}
${\bullet}$ $b=2$. We have 16 D9-branes. The orientifold projection is of the $Sp$ type.
The solution to the twisted tadpole cancellation conditions is given by:
\begin{equation}
 \gamma_{g,9}={\mbox{diag}}(\omega {\bf I}_2,\omega^{-1} {\bf I}_2, {\bf I}_4 )~.
\end{equation} 
The massless spectrum of this model is given in Table I. The gauge group is $U(4)\otimes Sp(8)$. The superpotential reads:
\begin{equation}
 {\cal W}=\epsilon_{s{s^\prime}s^{\prime\prime}} \Phi_s Q_{s^\prime} Q_{s^{\prime\prime}}~.
\end{equation}
${\bullet}$ $b=4$. We have 8 D9-branes. The orientifold projection is of the $SO$ type.
The solution to the twisted tadpole cancellation conditions is given by:
\begin{equation}
 \gamma_{g,9}={\mbox{diag}}(\omega {\bf I}_2,\omega^{-1} {\bf I}_2)~.
\end{equation} 
The massless spectrum of this model is given in Table I. The gauge group is $U(4)$. (This model
was briefly discussed in \cite{Sagnotti}.) There are no renormalizable couplings in this case.\\
${\bullet}$ $b=6$. We have 4 D9-branes. The orientifold projection is of the $Sp$ type.
The solution to the twisted tadpole cancellation conditions is given by:
\begin{equation}
 \gamma_{g,9}={\bf I}_2~.
\end{equation} 
The massless spectrum of this model is given in Table I. The gauge group is $Sp(4)$. There is no matter charged under the $Sp(4)$ gauge group in this model.

\subsection{The ${\bf Z}_7$ Orbifold}

{}Let $g$ be the generator of the orbifold group $\Gamma\approx{\bf Z}_7$. The action of $g$ on the
complex coordinates $z_s$ is given by:
\begin{equation}
 gz_1=\omega z_1~,~~~gz_2=\omega^2 z_2~,~~~g z_3=\omega^4 z_3~, ~~~
 \omega=\exp(2\pi i/7)~.
\end{equation} 
(The corresponding Calabi-Yau three-fold ${\cal M}=T^6/\Gamma$ has the following Hodge numbers: $(h^{1,1},h^{2,1})=(24,0)$. Thus, there are 24 chiral supermultiplets in the closed 
string sector.)
In this case we have D9-branes only.
The Chan-Paton matrices corresponding to the solutions of tadpole cancellation conditions depend on the rank $b$ of the $B_{ij}$ matrix.\\
${\bullet}$ $b=0$. We have 32 D9-branes. The orientifold projection is of the $SO$ type.
The solution to the twisted tadpole cancellation conditions is given by:
\begin{equation}
 \gamma_{g,9}={\mbox{diag}}(\omega {\bf I}_2,\omega^{2} {\bf I}_2, 
 \omega^3 {\bf I}_2,\omega^4 {\bf I}_2, \omega^5 {\bf I}_2, \omega^6 {\bf I}_2,
 {\bf I}_4 )~.
\end{equation} 
The massless spectrum of this model is given in Table I. The gauge group is $U(4)^3\otimes SO(8)$. (This model was originally constructed in \cite{KS1}.) The superpotential reads:
\begin{equation}
 {\cal W}=\epsilon_{s{s^\prime}s^{\prime\prime}} P_s P_{s^\prime} Q_{s^{\prime\prime}}+
 \epsilon_{s{s^\prime}s^{\prime\prime}} Q_s R_{s^\prime} \Phi_{s^{\prime\prime}}+
 \epsilon_{s{s^\prime}s^{\prime\prime}} R_s R_{s^\prime} R_{s^{\prime\prime}}~.
\end{equation}
${\bullet}$ $b=6$. We have 4 D9-branes. The orientifold projection is of the $SO$ type.
The solution to the twisted tadpole cancellation conditions is given by:
\begin{equation}
 \gamma_{g,9}={\bf I}_2~.
\end{equation} 
The massless spectrum of this model is given in Table I. The gauge group is $SO(4)$. There is no matter charged under the $SO(4)$ gauge group in this model.

\subsection{The ${\bf Z}_3\otimes {\bf Z}_3$ Orbifold}

{}Let $g_1$ and $g_2$ be the generators of the first and the second ${\bf Z}_3$ subgroups of the orbifold group $\Gamma\approx{\bf Z}_3\otimes {\bf Z}_3$. The action of $g_1$ and $g_2$ on the complex coordinates $z_s$ is given by ($\omega=\exp(2\pi i/3)$):
\begin{eqnarray}
 &&g_1 z_1=\omega z_1~,~~~g_1 z_2=\omega^{-1} z_2~,~~~g_1 z_3=z_3~,\\
 &&g_2 z_1=z_1~,~~~g_2 z_2=\omega  z_2~,~~~g_2 z_3=\omega^{-1} z_3~.
\end{eqnarray} 
(The corresponding Calabi-Yau three-fold ${\cal M}=T^6/\Gamma$ has the following Hodge numbers: $(h^{1,1},h^{2,1})=(84,0)$. Thus, there are 84 chiral supermultiplets in the closed 
string sector.)
In this case we have D9-branes only.
The Chan-Paton matrices corresponding to the solutions of tadpole cancellation conditions depend on the rank $b$ of the $B_{ij}$ matrix.\\
${\bullet}$ $b=0$. We have 32 D9-branes. The orientifold projection is of the $SO$ type.
The solution to the twisted tadpole cancellation conditions is given by:
\begin{eqnarray}
 &&\gamma_{g_1,9}={\mbox{diag}}(\omega {\bf I}_4,\omega^{-1} {\bf I}_4, {\bf I}_8 )~,\\
 &&\gamma_{g_2,9}={\mbox{diag}}(\omega^{-1} {\bf I}_2,{\bf I}_2,\omega {\bf I}_2,{\bf I}_2, 
 \omega {\bf I}_2,\omega^{-1} {\bf I}_2, {\bf I}_4 )~.
\end{eqnarray}
The massless spectrum of this model is given in Table I. The gauge group is $U(4)^3\otimes SO(8)$. (This model was originally constructed in \cite{KS2}.) The superpotential reads:
\begin{equation}
 {\cal W}=\epsilon_{s{s^\prime}s^{\prime\prime}} P_s P_{s^\prime} Q_{s^{\prime\prime}}~.
\end{equation}
${\bullet}$ $b=2$. We have 16 D9-branes. The orientifold projection is of the $Sp$ type.
The solution to the twisted tadpole cancellation conditions (up to equivalent representations) is given by:
\begin{eqnarray}
 &&\gamma_{g_1,9}={\mbox{diag}}(\omega {\bf I}_4,\omega^{-1} {\bf I}_4)~,\\
 &&\gamma_{g_2,9}={\mbox{diag}}(\omega {\bf I}_2,{\bf I}_2,\omega^{-1} {\bf I}_2,{\bf I}_2)~.
\end{eqnarray}
(Here we have chosen $B_{12}=1/2$, $B_{34}=B_{56}=0$.)
The massless spectrum of this model is given in Table I. The gauge group is $U(4)\otimes U(4)$. The superpotential reads:
\begin{equation}
 {\cal W}=Q R \Phi~.
\end{equation}
${\bullet}$ $b=4$. We have 8 D9-branes. The orientifold projection is of the $SO$ type.
The solution to the twisted tadpole cancellation conditions is given by:
\begin{eqnarray}
 &&\gamma_{g_1,9}={\bf I}_4~,\\
 &&\gamma_{g_2,9}={\mbox{diag}}(\omega {\bf I}_2,\omega^{-1} {\bf I}_2)~.
\end{eqnarray} 
(Here we have chosen $B_{12}=B_{34}=1/2$, $B_{56}=0$.)
The massless spectrum of this model is given in Table I. The gauge group is $U(4)$. There are no renormalizable couplings in this case.\\
${\bullet}$ $b=6$. We have 4 D9-branes. The orientifold projection is of the $Sp$ type.
The solution to the twisted tadpole cancellation conditions is given by:
\begin{equation}
 \gamma_{g_1,9}=\gamma_{g_2,9}={\bf I}_2~.
\end{equation} 
The massless spectrum of this model is given in Table I. The gauge group is $Sp(4)$. There is no matter charged under the $Sp(4)$ gauge group in this model.

{}Here we note that for this orbifold we can consider cases where the twists $g_1$ and/or $g_2$
are accompanied by ${\bf Z}_3$ shifts in the two-tori left invariant under the corresponding twist.
This changes the Hodge numbers of the corresponding orbifold (hence a different number of closed string sector chiral supermultiplets) but does not affect the open string sector.

\subsection{The ${\bf Z}_2\otimes {\bf Z}_2$ Orbifold}

{}Let $R_1$ and $R_2$ be the generators of the first and the second ${\bf Z}_2$ subgroups of the orbifold group $\Gamma=\{1,R_1,R_2,R_3\}\approx{\bf Z}_2\otimes {\bf Z}_2$. 
(Here $R_s R_{s^\prime}=R_{s^{\prime\prime}}$, $s\not=s^\prime\not=s^{\prime\prime}\not=s$.)
The action of $R_s$ on the complex coordinates $z_{s^\prime}$ is given by (there is no summation over the repeated indices here):
\begin{equation}
 R_s z_{s^\prime}=-(-1)^{\delta_{ss^\prime}} z_{s^\prime}~.
\end{equation}
(The corresponding Calabi-Yau three-fold ${\cal M}=T^6/\Gamma$ has the following Hodge numbers: $(h^{1,1},h^{2,1})=(51,3)$. Thus, there are 54 chiral supermultiplets in the closed 
string sector.)
In this case we have $32/2^{b/2}$ D9-branes, and three sets of D5-branes with 
$32/2^{b/2}$ D5-branes in each set. Thus, the world-volumes of the D$5_s$-branes are
the four non-compact space-time coordinates plus the two-torus parametrized by the complex coordinate $z_s$. Up to equivalent representations the Chan-Paton matrices are given by:
\begin{equation}
 \gamma_{R_s,9}=i\sigma_s\otimes {\bf I}_{8/2^{b/2}}~.
\end{equation}  
Here $\sigma_s$ are Pauli matrices. (The action on the D$5_s$-branes is similar.) The spectrum of this model is given in Table I. (The model with $b=0$ was originally constructed in \cite{BL}.)
The orientifold projection in the 99 sector is of the $SO$ type. The gauge group is $[Sp(16/2^{b/2})]_{99}\otimes\bigotimes_{s=1}^3 [Sp(16/2^{b/2})]_{5_s 5_s}$.  

{}As we already mentioned, the multiplicity of states in the $95_s$ and $5_s 5_{s^\prime}$ sectors is $k=2^{b/2}$. This multiplicity is labeled by $\alpha=1,\dots,k$. In the following it will be convenient to use a different basis for the index $\alpha$. The states in this sector carry 
charges under a discrete group $({\bf Z}_2)^{b/2}$ (which is the source of this multiplicity). We will use the following basis for $\alpha$. Let $\alpha$ be a vector with $b/2$ entries:
$\alpha=(\alpha_1,\dots,\alpha_{b/2})$, where the components $\alpha_A=\pm1$. Let us define
the dot product of two such vectors $\alpha$ and $\beta$ as $\alpha\cdot\beta=
(\alpha_1\beta_1,\dots,\alpha_{b/2}\beta_{b/2})$. Let us also introduce the following notation:
\begin{eqnarray}
 && {\cal Y}_{\alpha\beta}=1~,~~~ \alpha\cdot\beta=(+1,\dots,+1)~,\\
 && {\cal Y}_{\alpha\beta}=0~,~~~{\mbox{otherwise}}~,\\
 && {\cal Y}_{\alpha\beta\gamma}=1~,~~~ \alpha\cdot\beta\cdot\gamma=(+1,\dots,+1)~,\\
 && {\cal Y}_{\alpha\beta\gamma}=0~,~~~{\mbox{otherwise}}~.
\end{eqnarray}
Using these notations, the superpotential can be written as (the $({\bf Z}_2)^{b/2}$ charges
must be conserved in the scattering):
\begin{eqnarray}
 {\cal W}=&&\epsilon_{r r^\prime r^{\prime\prime}}\Phi_r \Phi_{r^\prime} \Phi_{r^{\prime\prime}} + 
 \epsilon_{r r^\prime r^{\prime\prime}} \Phi_r^s \Phi_{r^\prime}^s \Phi_{r^{\prime\prime}}^s +
 \epsilon_{s s^\prime r} {\cal Y}_{\alpha\beta} \Phi^s_r Q^{\alpha s s^\prime} Q^{\beta 
 s s^\prime} +\nonumber\\
 &&{\cal Y}_{\alpha\beta} \Phi_s Q^{\alpha s} Q^{\beta s}+{\cal Y}_{\alpha\beta\gamma}
 Q^{\alpha s s^\prime} Q^{\beta s^\prime s^{\prime\prime}} Q^{\gamma s^{\prime\prime} s} +
 {\cal Y}_{\alpha\beta\gamma} Q^{\alpha s s^\prime} Q^{\beta s} Q^{\gamma s^\prime}~.
\end{eqnarray}

{}Just as in the ${\bf Z}_3\otimes {\bf Z}_3$ case, for the ${\bf Z}_2\otimes {\bf Z}_2$ orbifold we can consider cases where the $R_1$ and/or $R_2$ twists are accompanied by ${\bf Z}_2$ shifts in the two-tori left invariant under the corresponding twist.
This changes the Hodge numbers of the corresponding orbifold (hence a different number of closed string sector chiral supermultiplets). Thus, if only $R_1$ is accompanied by a shift
(in the first two-torus parametrized by $z_1$), then the Hodge numbers are $(h^{1,1},h^{2,1})=(19,19)$ (so that the total number of closed string chiral multiplets is 38).
If the $R_2$ twist is also accompanied by a shift (in the second two-torus parametrized by $z_2$), then the Hodge numbers are $(h^{1,1},h^{2,1})=(11,11)$ (so that the total number of closed string chiral multiplets is 22).
Unlike in the ${\bf Z}_3\otimes {\bf Z}_3$ case, however, the open string sector does get affected by the presence of shifts.\\ 
$\bullet$ $(h^{1,1},h^{2,1})=(19,19)$. We have only 
two different types of D5-branes, namely, D$5_2$- and D$5_3$-branes, as well as D9-branes. 
The massless spectrum of this model is given in Table I. The gauge group is $[Sp(16/2^{b/2})]_{99}\otimes[Sp(16/2^{b/2})]_{5_2 5_2}\otimes[Sp(16/2^{b/2})]_{5_3 5_3}$.
The superpotential reads:
\begin{eqnarray}
 {\cal W}=&&\epsilon_{r r^\prime r^{\prime\prime}}\Phi_r \Phi_{r^\prime} \Phi_{r^{\prime\prime}} + 
 \epsilon_{r r^\prime r^{\prime\prime}} \Phi_r^s \Phi_{r^\prime}^s \Phi_{r^{\prime\prime}}^s +
 \epsilon_{s s^\prime r} {\cal Y}_{\alpha\beta} \Phi^s_r Q^{\alpha s s^\prime} Q^{\beta 
 s s^\prime} +\nonumber\\
 &&{\cal Y}_{\alpha\beta} \Phi_s Q^{\alpha s} Q^{\beta s} +
 {\cal Y}_{\alpha\beta\gamma} Q^{\alpha s s^\prime} Q^{\beta s} Q^{\gamma s^\prime}~.
\end{eqnarray}
$\bullet$ $(h^{1,1},h^{2,1})=(11,11)$. We have only 
one type of D5-branes, namely, D$5_3$-branes, as well as D9-branes.
The massless spectrum of this model is given in Table I. The gauge group is $[Sp(16/2^{b/2})]_{99}\otimes[Sp(16/2^{b/2})]_{5_3 5_3}$.
The superpotential reads:
\begin{eqnarray}
 {\cal W}=\epsilon_{r r^\prime r^{\prime\prime}}\Phi_r \Phi_{r^\prime} \Phi_{r^{\prime\prime}} + 
 \epsilon_{r r^\prime r^{\prime\prime}} \Phi_r^\prime \Phi_{r^\prime}^\prime  
 \Phi_{r^{\prime\prime}}^\prime +
 {\cal Y}_{\alpha\beta} \Phi_3 Q^{\alpha } Q^{\beta }~. 
\end{eqnarray}

\subsection{The ${\bf Z}_6$ Orbifold}

{}Let $g$ and $R$ be the generators of the ${\bf Z}_3$ and ${\bf Z}_2$ subgroups of the orbifold
group $\Gamma\approx{\bf Z}_6\approx{\bf Z}_3\otimes{\bf Z}_2$. The action of $g$ and $R$ on the complex coordinates $z_s$ is given by:
\begin{eqnarray}
 &&g z_s=\omega z_s~,~~~\omega=\exp(2\pi i/3)~,\\
 &&R z_1=-z_1~,~~~R z_2=-  z_2~,~~~R z_3= z_3~.
\end{eqnarray} 
(The corresponding Calabi-Yau three-fold ${\cal M}=T^6/\Gamma$ has the following Hodge numbers: $(h^{1,1},h^{2,1})=(29,5)$. Thus, there are 34 chiral supermultiplets in the closed 
string sector.)
In this case we have $32/2^{b/2}$ D9-branes, and one set of $32/2^{b/2}$ D5-branes. The world-volumes of the D5-branes are
the four non-compact space-time coordinates plus the two-torus parametrized by the complex coordinate $z_3$. The Chan-Paton matrices corresponding to the solutions of tadpole cancellation conditions depend on the rank $b$ of the $B_{ij}$ matrix.\\
${\bullet}$ $b=0$. We have 32 D9-branes and 32 D5-branes. 
The orientifold projection in the 99 sector is of the $SO$ type.
The solution to the twisted tadpole cancellation conditions is given by:
\begin{eqnarray}
 &&\gamma_{g,9}=\gamma_{g,5}=
 {\mbox{diag}}(\omega {\bf I}_6,\omega^{-1} {\bf I}_6, {\bf I}_4 )~,\\
 &&\gamma_{R,9}=\gamma_{R,5}=
 {\mbox{diag}}(i {\bf I}_3,-i{\bf I}_3,i {\bf I}_3,-i{\bf I}_3, 
 i{\bf I}_2,-i {\bf I}_2 )~.
\end{eqnarray}
The massless spectrum of this model is given in Table II. The gauge group is $[U(6)^2\otimes U(4)]_{99}\otimes [U(6)^2\otimes U(4)]_{55}$. (This model was originally constructed in \cite{KS2}.) The superpotential reads:
\begin{eqnarray}
 {\cal W}=&&P_1 {\widetilde P}_2 R+ P_2 {\widetilde P}_1 R +
 \Phi_1 P_2 P_3 +\Phi_2 P_1 P_3 + {\widetilde \Phi}_1 {\widetilde P}_2 {\widetilde P}_3 +
 {\widetilde \Phi}_2 {\widetilde P}_1 {\widetilde P}_3+\nonumber\\
 &&
 P_1^\prime {\widetilde P}_2^\prime R^\prime + P_2^\prime {\widetilde P}_1^\prime R^\prime +
 \Phi_1^\prime P_2^\prime P_3^\prime +\Phi_2^\prime P_1^\prime P_3^\prime + 
 {\widetilde \Phi}_1^\prime {\widetilde P}_2^\prime {\widetilde P}_3^\prime +
 {\widetilde \Phi}_2^\prime {\widetilde P}_1^\prime {\widetilde P}_3^\prime+\nonumber\\
 && S{\widetilde U} P_3+U{\widetilde S} {\widetilde P}_3 +T {\widetilde T} R+\nonumber\\
 &&S{\widetilde T} P_3^\prime+T{\widetilde S} {\widetilde P}_3^\prime +
 U {\widetilde U} R^\prime~.
\end{eqnarray}
${\bullet}$ $b=2$. We have 16 D9-branes and 16 D5-branes. 
The orientifold projection in the 99 sector is of the $Sp$ type.
The solution to the twisted tadpole cancellation conditions (up to equivalent representations) is given by:
\begin{eqnarray}
 &&\gamma_{g,9}=\gamma_{g,5}=
 {\mbox{diag}}(\omega {\bf I}_2,\omega^{-1} {\bf I}_2, {\bf I}_4 )~,\\
 &&\gamma_{R,9}=\gamma_{R,5}=
 {\mbox{diag}}(i ,-i,i ,-i, 
 i{\bf I}_2,-i {\bf I}_2 )~.
\end{eqnarray}
The massless spectrum of this model is given in Table II. The gauge group is $[U(2)^2\otimes U(4)]_{99}\otimes [U(2)^2\otimes U(4)]_{55}$. (This model has been recently discussed in a phenomenological context in \cite{ST}.) The superpotential reads:
\begin{eqnarray}
 {\cal W}=&&P_1 {\widetilde P}_2 R+ P_2 {\widetilde P}_1 R +
 \Phi_1 P_2 P_3 +\Phi_2 P_1 P_3 + {\widetilde \Phi}_1 {\widetilde P}_2 {\widetilde P}_3 +
 {\widetilde \Phi}_2 {\widetilde P}_1 {\widetilde P}_3+\nonumber\\
 &&
 P_1^\prime {\widetilde P}_2^\prime R^\prime + P_2^\prime {\widetilde P}_1^\prime R^\prime +
 \Phi_1^\prime P_2^\prime P_3^\prime +\Phi_2^\prime P_1^\prime P_3^\prime + 
 {\widetilde \Phi}_1^\prime {\widetilde P}_2^\prime {\widetilde P}_3^\prime +
 {\widetilde \Phi}_2^\prime {\widetilde P}_1^\prime {\widetilde P}_3^\prime+\nonumber\\
 &&{\cal Y}_{\alpha\beta} S^\alpha {\widetilde U}^\beta P_3+
 {\cal Y}_{\alpha\beta}U^\alpha {\widetilde S}^\beta {\widetilde P}_3 +
 {\cal Y}_{\alpha\beta}T^\alpha {\widetilde T}^\beta R+\nonumber\\
 &&{\cal Y}_{\alpha\beta} S^\alpha {\widetilde T}^\beta  P_3^\prime+
 {\cal Y}_{\alpha\beta} T^\alpha {\widetilde S}^\beta {\widetilde P}_3^\prime +
 {\cal Y}_{\alpha\beta} U^\alpha  {\widetilde U}^\beta R^\prime~.
\end{eqnarray}
${\bullet}$ $b=4$. We have 8 D9-branes and 8 D5-branes. The orientifold projection in the 99 sector is of the $SO$ type.
The solution to the twisted tadpole cancellation conditions is given by:
\begin{eqnarray}
 &&\gamma_{g,9}=\gamma_{g,5}=
 {\mbox{diag}}(\omega {\bf I}_2,\omega^{-1} {\bf I}_2)~,\\
 &&\gamma_{R,9}=\gamma_{R,5}=
 {\mbox{diag}}(i ,-i,i ,-i)~.
\end{eqnarray} 
The massless spectrum of this model is given in Table III. The gauge group is $[U(2)^2]_{99}\otimes [U(2)^2]_{55}$. There are no renormalizable couplings in this case.\\
${\bullet}$ $b=6$. We have 4 D9-branes and 4 D5-branes. The orientifold projection in the 99 sector is of the $Sp$ type.
The solution to the twisted tadpole cancellation conditions is given by:
\begin{eqnarray}
 &&\gamma_{g,9}=\gamma_{g,5}={\bf I}_2~,\\
  &&\gamma_{R,9}=\gamma_{R,5}=
 {\mbox{diag}}(i ,-i)~.
\end{eqnarray}
The massless spectrum of this model is given in Table III. The gauge group is $[U(2)]_{99}\otimes [U(2)]_{55}$. There is no matter charged under the gauge group in this model.

\subsection{The ${\bf Z}_2\otimes {\bf Z}_2\otimes {\bf Z}_3$ Orbifold}

{}Let $g$, $R_1$ and $R_2$ be the generators of the ${\bf Z}_3$ and the two ${\bf Z}_2$ subgroups of the orbifold
group $\Gamma\approx{\bf Z}_2\otimes{\bf Z}_2\otimes {\bf Z}_3$. The action of $g$ and $R_s$ ($R_3=R_1R_2$) on the complex coordinates $z_{s^\prime}$ is given by (there is no summation over the repeated indices here):
\begin{eqnarray}
 &&g z_s=\omega z_s~,~~~\omega=\exp(2\pi i/3)~,\\
 &&R_s z_{s^\prime}=-(-1)^{\delta_{ss^\prime}} z_{s^\prime}~.
\end{eqnarray} 
(The corresponding Calabi-Yau three-fold ${\cal M}=T^6/\Gamma$ has the following Hodge numbers: $(h^{1,1},h^{2,1})=(36,0)$. Thus, there are 36 chiral supermultiplets in the closed 
string sector.)
As in the ${\bf Z}_2\otimes {\bf Z}_2$ case, here we have $32/2^{b/2}$ D9-branes, and three sets of D5-branes with $32/2^{b/2}$ D5-branes in each set. The world-volumes of the D$5_s$-branes are
the four non-compact space-time coordinates plus the two-torus parametrized by the complex coordinate $z_s$. The Chan-Paton matrices corresponding to the solutions of tadpole cancellation conditions depend on the rank $b$ of the $B_{ij}$ matrix.\\
${\bullet}$ $b=0$. We have 32 D9-branes and three sets of D$5_s$-branes with 32 D5-branes in each set. 
The orientifold projection in the 99 sector is of the $SO$ type.
The solution to the twisted tadpole cancellation conditions is given by:
\begin{eqnarray}
 &&\gamma_{g,9}={\mbox{diag}}({\bf W}\otimes {\bf I}_3, {\bf I}_4)~,\\
 &&\gamma_{R_s,9}=i\sigma_s \otimes {\bf I}_8~.
\end{eqnarray}
Here ${\bf W}={\mbox{diag}}(\omega,\omega,\omega^{-1},\omega^{-1})$. (The action on the D$5_s$-branes is similar.)
The massless spectrum of this model is given in Table III. The gauge group is $[U(6)\otimes Sp(4)]_{99}\otimes \bigotimes_{s=1}^3 [U(6)\otimes Sp(4)]_{5_s 5_s}$. 
(This model was originally constructed in \cite{zk}.) The superpotential reads:
\begin{eqnarray}
 {\cal W}=&&\epsilon_{r r^\prime r^{\prime\prime}} \Phi_r\chi_{r^\prime}\chi_{r^{\prime\prime}}
 +\epsilon_{r r^\prime r^{\prime\prime}} \Phi^s_r\chi^s_{r^\prime}\chi^s_{r^{\prime\prime}}
 +\epsilon_{s s^\prime r}
  \Phi^s_r Q^{s s^\prime} Q^{s s^\prime} + 
 \epsilon_{s s^\prime r}
  \chi^s_r P^{s s^\prime} R^{s s^\prime} +\nonumber\\
 && \Phi_s Q^{s} Q^{s} + \chi_s P^{s} R^{s} +P^{ss^\prime} Q^{s^\prime s^{\prime\prime}} 
 R^{s^{\prime\prime} s}+\nonumber\\
 \label{sup}
 &&Q^{ss^\prime} P^s Q^{s^\prime}+R^{ss^\prime} Q^s P^{s^\prime}+P^{ss^\prime} R^s  
 R ^{s^\prime}~.
\end{eqnarray}
${\bullet}$ $b=2$. We have 16 D9-branes and three sets of D$5_s$-branes with 16 D5-branes in each set. 
The orientifold projection in the 99 sector is of the $Sp$ type.
The solution to the twisted tadpole cancellation conditions is given by:
\begin{eqnarray}
 &&\gamma_{g,9}={\mbox{diag}}({\bf W}, {\bf I}_4)~,\\
 &&\gamma_{R_s,9}=i\sigma_s \otimes {\bf I}_4~.
\end{eqnarray}
(The action on the D$5_s$-branes is similar.)
The massless spectrum of this model is given in Table III. The gauge group is $[U(2)\otimes Sp(4)]_{99}\otimes \bigotimes_{s=1}^3 [U(2)\otimes Sp(4)]_{5_s 5_s}$. The superpotential reads:
\begin{eqnarray}
 {\cal W}=&&\epsilon_{r r^\prime r^{\prime\prime}} \Phi_r\chi_{r^\prime}\chi_{r^{\prime\prime}}
 +\epsilon_{r r^\prime r^{\prime\prime}} \Phi^s_r\chi^s_{r^\prime}\chi^s_{r^{\prime\prime}}
 +\epsilon_{s s^\prime r} {\cal Y}_{\alpha\beta}
  \Phi^s_r Q^{\alpha s s^\prime} Q^{\beta s s^\prime} + 
 \epsilon_{s s^\prime r} {\cal Y}_{\alpha\beta}
  \chi^s_r P^{\alpha s s^\prime} R^{\beta s s^\prime} +\nonumber\\
 && {\cal Y}_{\alpha\beta} \Phi_s Q^{\alpha s} Q^{\beta s} + 
 {\cal Y}_{\alpha\beta} \chi_s P^{\alpha s} R^{\beta s} +
 {\cal Y}_{\alpha\beta\gamma} P^{\alpha ss^\prime} Q^{\beta s^\prime s^{\prime\prime}} 
 R^{\gamma s^{\prime\prime} s}+\nonumber\\
 &&{\cal Y}_{\alpha\beta\gamma} Q^{\alpha ss^\prime} P^{\beta s} Q^{\gamma  
 s^\prime}+{\cal Y}_{\alpha\beta\gamma} R^{\alpha ss^\prime} Q^{\beta s} P^{\gamma 
 s^\prime}+{\cal Y}_{\alpha\beta\gamma} P^{\alpha ss^\prime} R^{\beta s} R^{\gamma 
 s^\prime}~.
\end{eqnarray}
${\bullet}$ $b=4$. We have 8 D9-branes and three sets of D$5_s$-branes with 8 D5-branes in each set. The orientifold projection in the 99 sector is of the $SO$ type.
The solution to the twisted tadpole cancellation conditions is given by:
\begin{eqnarray}
 &&\gamma_{g,9}={\bf W}~,\\
 &&\gamma_{R_s,9}=i\sigma_s \otimes {\bf I}_2~.
\end{eqnarray}
(The action on the D$5_s$-branes is similar.)
The massless spectrum of this model is given in Table III. The gauge group is $[U(2)]_{99}\otimes \bigotimes_{s=1}^3 [U(2)]_{5_s 5_s}$. There are no renormalizable couplings in this case.\\
${\bullet}$ $b=6$. We have 4 D9-branes and three sets of D$5_s$-branes with 4 D5-branes in each set. The orientifold projection in the 99 sector is of the $Sp$ type.
The solution to the twisted tadpole cancellation conditions is given by:
\begin{eqnarray}
 &&\gamma_{g,9}={\bf I}_2~,\\
 &&\gamma_{R_s,9}=i\sigma_s~.
\end{eqnarray}
(The action on the D$5_s$-branes is similar.)
The massless spectrum of this model is given in Table III. The gauge group is $[Sp(2)]_{99}\otimes \bigotimes_{s=1}^3 [Sp(2)]_{5_s 5_s}$. There is no matter charged under the gauge group in this model.

\subsection{Comments}

{}Here the following comments are in order. 
(These remarks apply to all the cases, including those discussed in the next section, 
except for the ${\bf Z}_2\otimes {\bf Z}_2$ models.) In all of the models we discuss in this paper
(except for the ${\bf Z}_2\otimes {\bf Z}_2$ cases) 
there is present at least one anomalous $U(1)$ \cite{anom} in the
massless open string spectrum. (More precisely, there are as many anomalous $U(1)$'s as
different types of D-branes. For instance, in the ${\bf Z}_2\otimes {\bf Z}_2\otimes {\bf Z}_3$
model there are 4 anomalous $U(1)$'s: one coming from the 99 open string sector, and the other three coming from the $5_s 5_s$ ($s=1,2,3$) open string sectors.) Presence of these anomalous $U(1)$'s
implies that there are corresponding Fayet-Iliopoulos D-terms which must be cancelled 
via a generalized Green-Schwarz mechanism \cite{GS}. 
The fields responsible for breaking these $U(1)$'s are some of the closed string sector singlets
(which transform
non-trivially under the anomalous $U(1)$ gauge transformations) corresponding to the
orbifold blow-up modes \cite{Sagnotti,ZK,KST}. On the other hand, as explained
at length in \cite{KST}, 
in all these cases the orbifold
singularities must be blown-up for the orientifold action to be consistent. In this process all
the non-perturbative (from the orientifold viewpoint) states naively expected in the massless
spectrum (see section II of this paper and \cite{KST} for more details) acquire masses via appropriate superpotentials \cite{ZK,KS1,KS2,KST}. That is, the massless spectra of these
models correspond to Type I compactifications on the appropriate {\em blown-up} orbifolds. 
(More precisely, all the orbifold singularities except for those in the ${\bf Z}_2$ twisted sectors must be blown up.)

{}Finally, we note that starting with the compact orientifolds considered in this paper, we can construct other Type I vacua by turning on {\em discrete} Wilson lines. The models with
Wilson lines are not difficult to construct. We will not discuss them in this paper, however.   

\section{The Non-Abelian Orbifold}\label{NA}

{}In the previous section we discussed perturbative $\Omega$ orientifolds of Type IIB on
$T^6/\Gamma$ where the orbifold group is Abelian. As pointed out in \cite{zura}, non-Abelian
orbifold groups generically lead to orientifolds which have no world-sheet description. Let us make this statement more precise. The point is that if the non-Abelian orbifold group $\Gamma$ (which for consistency we assume to act crystallographically on $T^6$) contains ${\bf Z}_2$ subgroups, then the corresponding orientifold 5-planes and D5-branes are always mutually non-local \cite{zura}. Such branes have no world-sheet description. (These states, however, 
can be appropriately described in the context of F-theory upon the corresponding T-duality transformation \cite{zura,KST}.) To avoid such mutually non-local orientifold 5-planes and D5-branes, we can search for non-Abelian orbifold groups which contain no ${\bf Z}_2$ subgroups. There is only one such orbifold group (that can act crystallographically on $T^6$)
to which we turn next.

\subsection{The Compact Model}

{}Let $\Gamma={\cal G}$, where ${\cal G}$ is a non-Abelian group generated by two elements
$g$ and $\theta$ whose action on the complex coordinates $z_s$ ($s=1,2,3$) is given by (here
for simplicity we can assume that the six-torus $T^6$ factorized into three two-tori, and the complex coordinates $z_s$ parametrize these two-tori):
\begin{eqnarray}
 &&g z_1=\omega z_1~,~~~g z_2=\omega^{-1} z_2~,~~~gz_3=z_3~,\\
 &&\theta z_1=z_2~,~~~\theta z_2=z_3~,~~~\theta z_3=z_1~,
\end{eqnarray}   
where $\omega=\exp(2\pi i/3)$.
The dimension of this non-Abelian group\footnote{The group ${\cal G}$ is a semidirect product of 
${\bf Z}_3$ and ${\bf Z}_3\otimes {\bf Z}_3$. It is the $n=3$ member of the infinite series
referred to as $\Delta(3n^2)$. (The non-Abelian group $\Delta(3n^2)$ is a semidirect product of
${\bf Z}_3$ and ${\bf Z}_n\otimes {\bf Z}_n$.)} is $|{\cal G}|=27$. It is not difficult to see that it contains
a ${\bf Z}_3\otimes {\bf Z}_3$ subgroup whose action on $z_s$ is the same as in the previous section. The element $\theta$ permutes the three two-tori (and this action does not commute with 
that of the generators of the above ${\bf Z}_3\otimes {\bf Z}_3$ subgroup). The Hodge numbers of the Calabi-Yau three-fold $T^6/{\cal G}$ are $(h^{1,1},h^{2,1})=(36,0)$. Thus, there are 36 chiral supermultiplets in the closed string sector. In this model we have D9-branes only. Since we must have permutational symmetry for the three two-tori, there are only two allowed values for the rank of the $B_{ij}$ matrix: $b=0$ and $b=6$. Let us consider these two cases separately.\\
$\bullet$ $b=0$. We have 32 D9-branes. The orientifold projection is of the $SO$ type. The solution to the twisted tadpole cancellation conditions is given by:
\begin{eqnarray}
 && \gamma_{g,9}={\mbox{diag}}({\bf X},{\bf X},{\bf I}_4)~,\\
 &&\gamma_{\theta,9}={\mbox{diag}}({\bf Y},{\bf Y}^{-1},{\bf I}_4)~.
\end{eqnarray} 
Here ${\bf X}={\mbox{diag}}({\bf I}_2,\omega{\bf I}_2,\omega^{-1}{\bf I}_2)$, and ${\bf Y}$
is a $6\times 6$ matrix that cyclically permutes the $2\times 2$ blocks in the matrix ${\bf X}$. (Thus, ${\bf Y}{\bf X}{\bf Y}^{-1}={\mbox{diag}}(\omega^{-1}{\bf I}_2, {\bf I}_2,\omega{\bf I}_2)$.)
The massless spectrum of this model is given in Table IV. The gauge group is $U(4)\otimes SO(8)$. The superpotential reads:
\begin{equation}
 {\cal W}=PP\Phi~.
\end{equation} 
$\bullet$ $b=6$. We have 4 D9-branes. The orientifold projection is of the $Sp$ type. The solution to the twisted tadpole cancellation conditions is given by:
\begin{equation}
 \gamma_{g,9}=\gamma_{\theta,9}={\bf I}_2~.
\end{equation} 
The massless spectrum of this model is given in Table IV. The gauge group is $Sp(4)$. There is no matter charged under the $Sp(4)$ gauge group in this model.

\subsection{The Non-Compact Model}

{}Since the $\Omega$ orientifold of Type IIB on $T^6/{\cal G}$ 
(which is equivalent to Type I compactified on the Calabi-Yau three-fold ${\cal M}=T^6/{\cal G}$)
discussed in the previous subsection contains D9-branes only, its heterotic dual must be perturbative. It is not difficult to construct this heterotic vacuum and explicitly verify 
along the lines of \cite{ZK,KS1,KS2} that the states in the twisted sectors of the heterotic model
(which are non-perturbative from the orientifold viewpoint) decouple once the orbifold singularities are blown up (which is in complete parallel with the corresponding discussions in \cite{ZK,KS1,KS2}). However, we will not give the details of this calculation here. (The matching
of the Type I and heterotic spectra requires calculating the superpotential for the twisted sector states for the heterotic model. This calculation is completely straightforward but the details are tedious to discuss due to the non-Abelian character of the orbifold.) Instead, we will explicitly
perform a different check along the lines devised in \cite{zura1}. We will 
consider\footnote{Here we point out that although naively one might expect the $\Omega J$ orientifolds of Type IIB on ${\bf C}^3/\Delta(3n^2)$ to be perturbatively consistent for all $n$, this is not the case. Thus, in \cite{zura1} it was shown that the $\Omega J$ orientifolds of Type IIB on ${\bf C}^3/{\bf Z}_n\otimes {\bf Z}_n$ are perturbatively consistent only for $n=3$.
In all the other cases there are non-perturbative states contributing to the corresponding massless spectra.} the $\Omega J$ orientifold of Type IIB on ${\bf C}^3/{\cal G}$ where $Jz_s=-z_s$ ($z_s$ being the three complex coordinates parametrizing ${\bf C}^3$). As we will see in a moment, this non-compact model is perturbatively consistent, that is, it has well defined world-sheet description (once the
orbifold singularities are blown up). As argued in \cite{zura1}\footnote{In particular, in \cite{zura1} it was found that there is one-to-one correspondence between the perturbative $\Omega$ orientifolds of Type IIB on $T^6/\Gamma$ and perturbative 
$\Omega J$ orientifolds of Type IIB on ${\bf C}^3/\Gamma$ in the sense that 
if the former is not expected to be perturbative, then the perturbative tadpole cancellation conditions have no solution for the former, and {\em visa-versa}.}, this provides a robust test for
the perturbative consistency of the corresponding $\Omega$ orientifold of Type IIB on $T^6/{\cal G}$ as the appearance of non-perturbative states in a given orbifold is a local phenomenon (that depends on the properties of D-branes which may or may not wrap (collapsed) two-cycles in the orbifold).

{}Note that in the $\Omega J$ orientifold of Type IIB on ${\bf C}^3/{\cal G}$ we have an orientifold 3-plane (but no orientifold 7-planes). The number of D3-branes we can introduce into this background is not constrained (unlike in the compact cases) by the untwisted tadpole cancellation conditions which is due to the fact that the space transverse to the D3-branes is non-compact (and the corresponding R-R flux can go to infinity). 
Also, both $SO$ and $Sp$ orientifold projections are allowed. (We will label these cases by
$\eta=-1$ and $\eta=+1$, respectively.)  
The solution to the twisted tadpole cancellation conditions is given by:
\begin{eqnarray}
 && \gamma_{g,9}={\mbox{diag}}({\bf X}_{3N-2\eta},{\bf X}_{3N-2\eta},
 {\bf U})~,\\
 &&\gamma_{\theta,9}={\mbox{diag}}({\bf Y}_{3N-2\eta},{\bf Y}_{3N-2\eta}^{-1},{\bf U}^\prime)~.
\end{eqnarray}
Here (unlike in all the previous cases) we have chosen to count D-brane images under the
orientifold action\footnote{This is done for the reason that the same convention was adapted in \cite{zura,zura1} where $\Omega J$ orientifolds of Type IIB on ${\bf C}^3/\Gamma$ were discussed for Abelian orbifold groups $\Gamma$.}. We are using the following notations:
\begin{equation}
{\bf X}_{3N-2\eta}={\mbox{diag}}({\bf I}_{3N-2\eta},\omega{\bf I}_{3N-2\eta},
\omega^{-1} {\bf I}_{3N-2\eta})~. 
\end{equation}
Furthermore, here ${\bf Y}_{3N-2\eta}$ is a $3(3N-2)\eta\otimes 3(3N-2\eta)$
matrix that cyclically permutes the $(3N-2\eta)\otimes (3N-2\eta)$ blocks in ${\bf X}_{3N-2\eta}$.
(Thus, ${\bf Y}_{3N-2\eta} {\bf X}_{3N-2\eta} {\bf Y}_{3N-2\eta}^{-1}=
{\mbox{diag}}(\omega^{-1} {\bf I}_{3N-2\eta},{\bf I}_{3N-2\eta},\omega{\bf I}_{3N-2\eta})$.) Finally,
\begin{eqnarray}
 && {\bf U}={\mbox{diag}}(\omega {\bf I}_{3N},\omega^{-1} {\bf I}_{3N},{\bf I}_{3N-2\eta})~,\\
 &&{\bf U}^\prime={\mbox{diag}}(\omega {\bf I}_{N},\omega^{-1} {\bf I}_{N},{\bf I}_{N},
 \omega^{-1} {\bf I}_{N},\omega {\bf I}_{N},{\bf I}_{N}, 
 \omega {\bf I}_{N},\omega^{-1} {\bf I}_{N},{\bf I}_{N-2\eta})~.
\end{eqnarray}
The number of D3-branes is $n_3=27N-14\eta$. The gauge group in this model is
$U(3N-2\eta)\otimes U(N)^4\otimes G_\eta (N-2\eta)$. Here $G_\eta=SO$ for $\eta=-1$ and $G_\eta=Sp$ for $\eta=+1$. The massless open string spectrum is given in Table IV. Note that the non-Abelian gauge anomaly is cancelled in this model. The number of {\em twisted} 
closed string sector 
chiral multiplets (neutral under the Chan-Paton gauge group) is $9$ in this model. The superpotential reads ($\ell=1,2,3,4$):
\begin{equation}
 {\cal W}=\chi\chi R +P_\ell {\widetilde P}_\ell R +P_\ell {\widetilde P}_\ell Q~.
\end{equation}

{}Note that the one-loop $\beta$-function coefficients $b_0(3N-2\eta)$, $b_0(N)$ and
$b_0(N-2\eta)$ for the $SU(3N-2\eta)$, $SU(N)$ and $G_\eta(N-2\eta)$ subgroups are independent of $N$:
\begin{eqnarray}
 &&b_0(3N-2\eta)=-3\eta~,\\
 &&b_0(N)=+2\eta~,\\
 &&b_0(N-2\eta)=+2\eta~.
\end{eqnarray}
Moreover, since this orientifold model is perturbative (and the world-sheet expansion is
well defined), following \cite{zura} we conclude that in the large $N$ limit of 't Hooft \cite{thooft} (defined as $N\rightarrow\infty$, $\lambda_s\rightarrow 0$, $\lambda_s N={\mbox{fixed}}$ \cite{BKV}, where $\lambda_s=g_{YM}^2$ is the Type IIB string coupling, and $g_{YM}$ is the
gauge coupling of the gauge theory living in the D3-branes) computation of any $M$-point correlation function in this model reduces to the corresponding computation in the parent {\em oriented} (in this case $U(27N-14\eta)$) gauge theory with ${\cal N}=4$ supersymmetry. In particular, in this limit the gauge coupling running is subleading ({\em i.e.}, it is suppressed by
powers of $1/N$). Along with six Abelian models constructed in \cite{zura,zura1}, the model of this section completes the construction of perturbative ${\cal N}=1$ 
$\Omega J$ orientifolds of Type IIB on
${\bf C}^3/\Gamma$ non-compact orbifolds (all of which possess the above mentioned properties in the large $N$ limit)\footnote{The ${\cal N}=2$ large $N$ gauge theories from orientifolds were discussed in detail in \cite{zura}, and also in \cite{FS}.}.

\section{Conclusions}

{}In this paper we have studied four dimensional ${\cal N}=1$ 
perturbative Type I vacua obtained by compactification on
toroidal $T^6/\Gamma$ orbifolds (with or without the NS-NS $B$-field). The number of such models is rather constrained as for most of the choices of $\Gamma$ there are non-perturbative states arising from D-branes wrapping various (collapsed) two-cycles in the orbifold. Our exhaustive analyses of these perturbative orientifolds completes the program of constructing
and understanding such compactifications. To go beyond this relatively limited set of vacua, one would need to better understand non-perturbative Type I and heterotic compactifications.
F-theory should provide, we believe, a very important complementary picture in moving along these lines (as it has proven to be the case in \cite{KST}). It is therefore also important to understand F-theory compactifications on Calabi-Yau four-folds (and, in particular, on toroidal
orbifolds $T^8/\Gamma$ with $SU(4)$ holonomy). Developing such tools is most likely going to be rather complicated. Nonetheless, a clever use of various dualities should be very helpful and facilitate the necessary analyses.

{}A bit more immediate direction for future research which appears to be rather interesting is 
understanding perturbative (from the orientifold viewpoint) non-supersymmetric Type I compactifications free of tadpoles and tachyons. From our experience with ${\cal N}=1$ models, we can expect that the number of such vacua should be rather limited (which makes it possible
to explore them in detail). One important implication of such a development would be construction of non-compact non-supersymmetric $\Omega J$ orientifolds of Type IIB on ${\bf C}^3/\Gamma$ (which would contain D3- and (possibly) D7-branes). This would provide
the first examples of non-supersymmetric gauge theories from orientifolds that in the large $N$ limit possess the same nice properties as their ${\cal N}=1$ counterparts discussed in
\cite{zura,zura1} and this paper. (Note that in the cases without orientifold planes there are infinitely many such non-supersymmetric models \cite{KaSi,LNV,BKV,BJ}.)

{}Finally, we briefly remark on phenomenological implications of the models discussed in this paper. One of these models, namely, the ${\bf Z}_2\otimes {\bf Z}_2\otimes {\bf Z}_3$ orbifold model of \cite{zk}, has an $SU(6)$ subgroup and three chiral families of $SU(6)$. 
However, as was already pointed out in \cite{zk}, 
in this model it appears to be impossible to break the $SU(6)$ gauge subgroup
down to that of the Standard Model (which is due to a peculiar matter content). Clearly, a better understanding of more generic Type I compactifications is more than desirable.

\acknowledgments

{}This work was supported in part by the grant NSF PHY-96-02074, 
and the DOE 1994 OJI award. I would like to thank Gary Shiu and Henry Tye for discussions. 
I would also like to thank Albert and Ribena Yu for 
financial support.

\newpage
\begin{figure}[t]
\epsfxsize=16 cm
\epsfbox{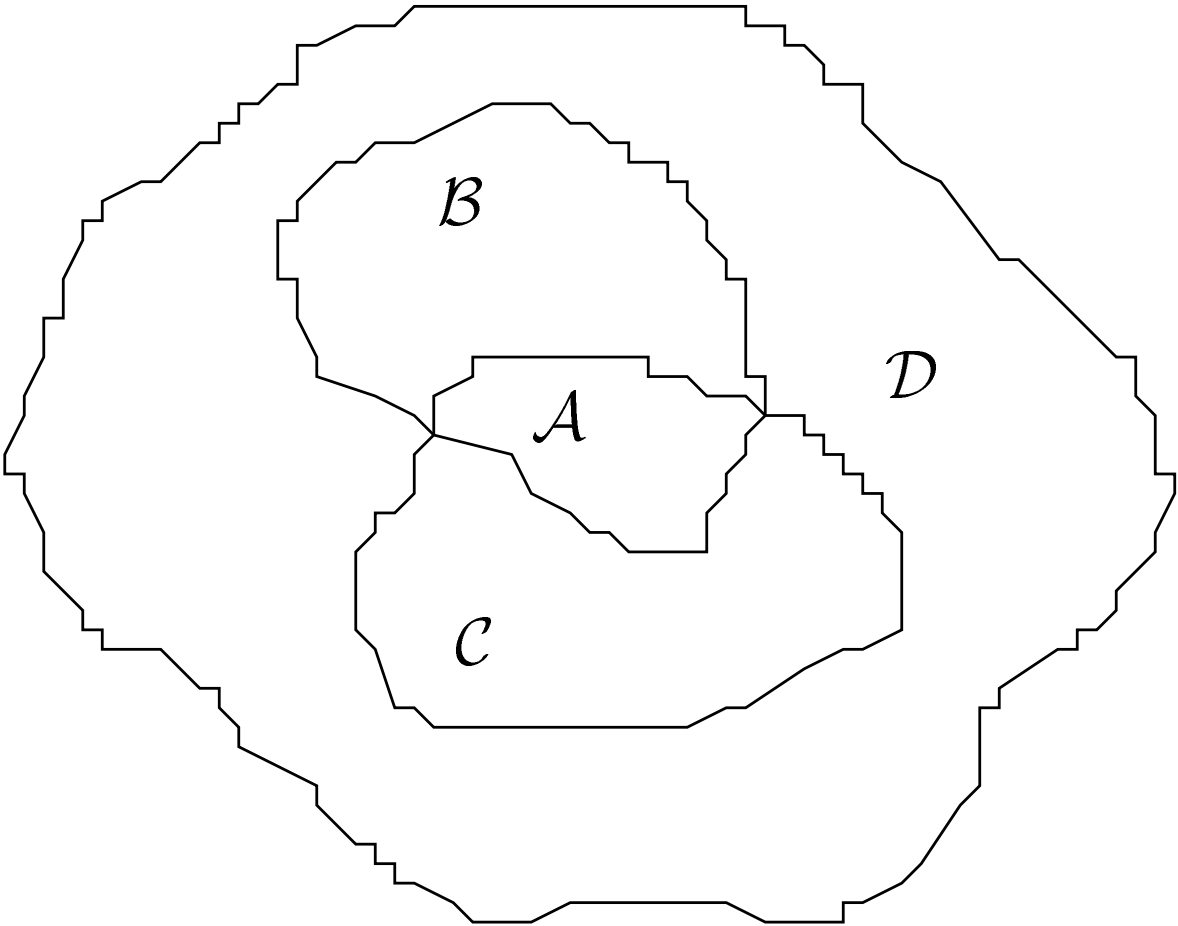}
\bigskip
\caption{A schematic picture of the space of four dimensional ${\cal N}=1$ Type I and
heterotic vacua. The region ${\cal A}\cup{\cal B}$ corresponds to perturbative Type I vacua.
The region ${\cal A}\cup{\cal C}$ corresponds to perturbative heterotic vacua. The vacua
in the region ${\cal A}$ are perturbative from both the Type I and heterotic viewpoints. The region ${\cal D}$ contains both non-perturbative Type I and heterotic vacua.}
\end{figure}

\begin{table}[t]
\begin{tabular}{|c|c|c|c|l|}
 Model & $b$ &Gauge Group  & Field & Charged Matter  
  \\
 \hline
${\bf Z}_3$ & 0 & $U(12)\otimes SO(8)$  & $\Phi_s$ &
 $3\times ({\bf 66},{\bf 1})(+2)_L$
 \\
            &&    &  $Q_s$ & $3\times ({\overline {\bf 12}},{\bf 8}_v)(-1)_L$  \\
\hline
 & 2 & $U(4)\otimes Sp(8)$  & $\Phi_s$ &
 $3\times ({\bf 10},{\bf 1})(+2)_L$
 \\
            &&    &  $Q_s$ & $3\times ({\overline {\bf 4}},{\bf 8})(-1)_L$  \\
\hline
  & 4 & $U(4)$  & $\Phi_s$ &
 $3\times {\bf 6}(+2)_L$ \\
\hline
  & 6 & $Sp(4)$  & &
 none \\
\hline
${\bf Z}_7$ & 0& $U(4)^3\otimes SO(8)$  &$ P_1$&
 $({\bf 4},{\bf 1},{\bf 1},{\bf 8}_v)(+1,0,0)_L$
  \\
  &&& $Q_1$ & $({\bf1},{\overline {\bf 4}},{\overline {\bf 4}},{\bf 1})(0,-1,-1)_L$ \\
    &&& $R_1$ & $({\overline {\bf 4}},{\bf 4},{\bf 1},{\bf 1})(-1,+1,0)_L$\\
    &&& $\Phi_1$ & $({\bf 1},{\bf 1},{\bf 6},{\bf 1})(0,0,+2)_L$\\
 & & & $P_{2,3},Q_{2,3},R_{2,3},\Phi_{2,3}$&plus cyclic permutations of the \\
 &&& & $U(4)\otimes U(4)\otimes U(4)$ irreps\\  
\hline
 & 6& $SO(4)$  & & none\\ 
\hline
${\bf Z}_3\otimes{\bf Z}_3$ & 0& $U(4)^3\otimes SO(8)$  &$ P_1$&
 $({\bf 4},{\bf 1},{\bf 1},{\bf 8}_v)(+1,0,0)_L$
  \\
  &&& $Q_1$ & $({\bf1},{\overline {\bf 4}},{\overline {\bf 4}},{\bf 1})(0,-1,-1)_L$ \\
    &&& $\Phi_1$ & $({\bf 6},{\bf 1},{\bf 1},{\bf 1})(-2,0,0)_L$\\
 & & & $P_{2,3},Q_{2,3},\Phi_{2,3}$ &plus cyclic permutations of the \\
 &&& & $U(4)\otimes U(4)\otimes U(4)$ irreps\\  
\hline
& 2& $U(4)\otimes U(4)$  &$ Q$&
 $({\overline {\bf 4}},{\overline {\bf 4}})(-1,-1)_L$
  \\
    &&& $R$ & $({\bf 4},{\overline {\bf 4}})(+1,-1)_L$\\
    &&& $\Phi$ & $({\bf 1},{\bf 10})(0,+2)_L$\\
\hline
& 4& $U(4)$  &$ \Phi$& ${\bf 6}(+2)_L$\\
\hline
& 6& $Sp(4)$ & &none\\
\hline
${\bf Z}_2\otimes {\bf Z}_2$ & $b$ & $[Sp(N)]_{99}\otimes$  & $\Phi_r$ &
 $3\times [{\bf A}_L]_{99}$
 \\
  (51,3)          &&  $\bigotimes_{s=1}^3 [Sp(N)]_{5_s5_s}$ 
 &  $\Phi_r^s$ & $3\times [{\bf A}_L]_{5_s 5_s}$  \\
            &&   
 &  $Q^{\alpha s}$ & $k \times [({\bf N};{\bf N}_s)_L]_{9 5_s}$  \\
            &  &   
 &  $Q^{\alpha ss^\prime}$ & $k \times [({\bf N}_s ;{\bf N}_{s^\prime})_L]_{ 5_s 5_{s^\prime}}$  \\
\hline
${\bf Z}_2\otimes {\bf Z}_2$ & $b$ & $[Sp(N)]_{99}\otimes$  & $\Phi_r$ &
 $3\times [{\bf A}_L]_{99}$
 \\
     (19,19)       &&  $\bigotimes_{s=2}^3 [Sp(N)]_{5_s5_s}$ 
 &  $\Phi_r^s$ & $3\times [{\bf A}_L]_{5_s 5_s}$  \\
            &&   
 &  $Q^{\alpha s}$ & $k \times [({\bf N};{\bf N}_s)_L]_{9 5_s}$  \\
            &  &   
 &  $Q^{\alpha ss^\prime}$ & $k \times [({\bf N}_s ;{\bf N}_{s^\prime})_L]_{ 5_s 5_{s^\prime}}$  \\
\hline
${\bf Z}_2\otimes {\bf Z}_2$ & $b$ & $[Sp(N)]_{99}\otimes[Sp(N)]_{5_35_3}$  & $\Phi_r$ &
 $3\times [{\bf A}_L]_{99}$
 \\
     (11,11)       &&    
 &  $\Phi_r^\prime$ & $3\times [{\bf A}_L]_{5_3 5_3}$  \\
            &&   
 &  $Q^{\alpha}$ & $k \times [({\bf N};{\bf N}_3)_L]_{9 5_3}$  \\
\hline
\end{tabular}
\caption{The massless open string spectra of the ${\cal N}=1$ orientifolds of Type IIB on $T^6/
{\bf Z}_3$, $T^6/ {\bf Z}_7$, $T^6/{\bf Z}_3\otimes {\bf Z}_3$  and $T^6/{\bf Z}_2\otimes {\bf Z}_2$.
The $U(1)$ charges are
given in parentheses. In the ${\bf Z}_2\otimes {\bf Z}_2$ case we have  $\alpha=1,\dots,k$, $k=2^{b/2}$, $N=16/k$,
and ${\bf A}$ stands for the two-index antisymmetric (reducible) representation of $Sp(N)$.}
\label{Z3} 
\end{table}

\begin{table}[t]
\begin{tabular}{|c|c|c|c|l|}
 Model & $b$ &Gauge Group  & Field & Charged Matter  
  \\
 \hline
${\bf Z}_6$ & 0 & $[U(6)\otimes U(6)\otimes U(4)]_{99}\otimes$  &  
 $\Phi_{1,2}$ &
 $2\times [({\bf 15},{\bf 1},{\bf 1};{\bf 1},{\bf 1},{\bf 1})(+2,0,0;0,0,0)_L]_{99}$
 \\
            &&  $[U(6)\otimes U(6)\otimes U(4)]_{55}$ 
 &  ${\widetilde \Phi}_{1,2}$ &
 $2\times [({\bf 1},{\overline {\bf 15}},{\bf 1};{\bf 1},{\bf 1},{\bf 1})(0,-2,0;0,0,0)_L]_{99}$  \\
            && 
 &  $P_{1,2}$ &
 $2\times [({\overline {\bf 6}},{\bf 1},{\overline {\bf 4}};
     {\bf 1},{\bf 1},{\bf 1})(-1,0,-1;0,0,0)_L]_{99}$  \\
          && 
 &  ${\widetilde P}_{1,2}$ &
 $2\times [({\bf 1},{\bf 6},{\bf 4};
     {\bf 1},{\bf 1},{\bf 1})(0,+1,+1;0,0,0)_L]_{99}$  \\
 &&&  $P_3$ &
  $[({\overline {\bf 6}},{\bf 1},{\bf 4};
     {\bf 1},{\bf 1},{\bf 1})(-1,0,+1;0,0,0)_L]_{99}$  \\
          && 
 &  ${\widetilde P}_3$ &
 $[({\bf 1},{\bf 6},{\overline {\bf 4}};
     {\bf 1},{\bf 1},{\bf 1})(0,+1,-1;0,0,0)_L]_{99}$  \\
 &&& $R$ &$ [({\bf 6},{\overline {\bf 6}},{\bf 1};
     {\bf 1},{\bf 1},{\bf 1})(+1,-1,0;0,0,0)_L]_{99}$  \\
 & &   &  
 $\Phi^\prime_{1,2}$ &
 $2\times [({\bf 1},{\bf 1},{\bf 1};{\bf 15},{\bf 1},{\bf 1})(0,0,0;+2,0,0)_L]_{55}$
 \\
            &&   
 &  ${\widetilde \Phi}^\prime_{1,2}$ &
 $2\times [({\bf 1},{\bf 1},{\bf 1};{\bf 1},{\overline {\bf 15}},{\bf 1})(0,0,0;0,-2,0)_L]_{55}$  \\
            && 
 &  $P^\prime_{1,2}$ &
 $2\times [(
     {\bf 1},{\bf 1},{\bf 1};{\overline {\bf 6}},{\bf 1},{\overline {\bf 4}})(0,0,0;-1,0,-1)_L]_{55}$  \\
          && 
 &  ${\widetilde P}^\prime_{1,2}$ &
 $2\times [(
     {\bf 1},{\bf 1},{\bf 1};{\bf 1},{\bf 6},{\bf 4})(0,0,0;0,+1,+1)_L]_{55}$  \\
 &&&  $P^\prime_3$ &
 $[(
     {\bf 1},{\bf 1},{\bf 1};{\overline {\bf 6}},{\bf 1},{\bf 4})(0,0,0;-1,0,+1)_L]_{55}$  \\
          && 
 &  ${\widetilde P}^\prime_3$ &
 $[(
     {\bf 1},{\bf 1},{\bf 1};{\bf 1},{\bf 6},{\overline {\bf 4}})(0,0,0;0,+1,-1)_L]_{55}$  \\
 &&& $R^\prime$ &$  [(
     {\bf 1},{\bf 1},{\bf 1};{\bf 6},{\overline {\bf 6}},{\bf 1})(0,0,0;+1,-1,0)_L]_{55}$  \\
&&& $S$         & $[({\bf 6},{\bf 1},{\bf 1};{\bf 6},{\bf 1},{\bf 1})
(+1,0,0;+1,0,0)_L]_{95}$\\
    &&& $T$       & $[({\bf 1},{\bf 6},{\bf 1};{\bf 1},{\bf 1},{\bf 4})
(0,+1,0;0,0,+1)_L]_{95}$ \\
&&& $U$ & $[({\bf 1},{\bf 1},{\bf 4};{\bf 1},{\bf 6},{\bf 1})
(0,0,+1;0,+1,0)_L]_{95}$\\
 &&& ${\widetilde S}$         & $[({\bf 1},\overline{\bf 6},{\bf 1};{\bf 1},\overline{\bf 6},
{\bf 1})(0,-1,0;0,-1,0)_L]_{95}$  \\ 
     &&&${\widetilde T}$      & $[(\overline{\bf 6},{\bf 1},{\bf 1};{\bf 1},{\bf 1},
\overline{\bf 4})(-1,0,0;0,0,-1)_L]_{95}$\\ 
&&& ${\widetilde U}$ & $[({\bf 1},{\bf 1},{\overline {\bf 4}};{\overline {\bf 6}},{\bf 1},{\bf 1})
(0,0,-1;-1,0,0)_L]_{95}$\\
\hline
 & 2 & $[U(2)\otimes U(2)\otimes U(4)]_{99}\otimes$  &  
 $\Phi_{1,2}$ &
 $2\times [({\bf 3},{\bf 1},{\bf 1};{\bf 1},{\bf 1},{\bf 1})(+2,0,0;0,0,0)_L]_{99}$
 \\
            &&  $[U(2)\otimes U(2)\otimes U(4)]_{55}$ 
 &  ${\widetilde \Phi}_{1,2}$ &
 $2\times [({\bf 1},{{\bf 3}},{\bf 1};{\bf 1},{\bf 1},{\bf 1})(0,-2,0;0,0,0)_L]_{99}$  \\
            && 
 &  $P_{1,2}$ &
 $2\times [({{\bf 2}},{\bf 1},{\overline {\bf 4}};
     {\bf 1},{\bf 1},{\bf 1})(-1,0,-1;0,0,0)_L]_{99}$  \\
          && 
 &  ${\widetilde P}_{1,2}$ &
 $2\times [({\bf 1},{\bf 2},{\bf 4};
     {\bf 1},{\bf 1},{\bf 1})(0,+1,+1;0,0,0)_L]_{99}$  \\
 &&&  $P_3$ &
  $[({{\bf 2}},{\bf 1},{\bf 4};
     {\bf 1},{\bf 1},{\bf 1})(-1,0,+1;0,0,0)_L]_{99}$  \\
          && 
 &  ${\widetilde P}_3$ &
 $[({\bf 1},{\bf 2},{\overline {\bf 4}};
     {\bf 1},{\bf 1},{\bf 1})(0,+1,-1;0,0,0)_L]_{99}$  \\
 &&& $R$ &$ [({\bf 2},{ {\bf 2}},{\bf 1};
     {\bf 1},{\bf 1},{\bf 1})(+1,-1,0;0,0,0)_L]_{99}$  \\
 & &   &  
 $\Phi^\prime_{1,2}$ &
 $2\times [({\bf 1},{\bf 1},{\bf 1};{\bf 3},{\bf 1},{\bf 1})(0,0,0;+2,0,0)_L]_{55}$
 \\
            &&   
 &  ${\widetilde \Phi}^\prime_{1,2}$ &
 $2\times [({\bf 1},{\bf 1},{\bf 1};{\bf 1},{{\bf 3}},{\bf 1})(0,0,0;0,-2,0)_L]_{55}$  \\
            && 
 &  $P^\prime_{1,2}$ &
 $2\times [(
     {\bf 1},{\bf 1},{\bf 1};{{\bf 2}},{\bf 1},{\overline {\bf 4}})(0,0,0;-1,0,-1)_L]_{55}$  \\
          && 
 &  ${\widetilde P}^\prime_{1,2}$ &
 $2\times [(
     {\bf 1},{\bf 1},{\bf 1};{\bf 1},{\bf 2},{\bf 4})(0,0,0;0,+1,+1)_L]_{55}$  \\
 &&&  $P^\prime_3$ &
 $[(
     {\bf 1},{\bf 1},{\bf 1};{{\bf 2}},{\bf 1},{\bf 4})(0,0,0;-1,0,+1)_L]_{55}$  \\
          && 
 &  ${\widetilde P}^\prime_3$ &
 $[(
     {\bf 1},{\bf 1},{\bf 1};{\bf 1},{\bf 2},{\overline {\bf 4}})(0,0,0;0,+1,-1)_L]_{55}$  \\
 &&& $R^\prime$ &$  [(
     {\bf 1},{\bf 1},{\bf 1};{\bf 2},{{\bf 2}},{\bf 1})(0,0,0;+1,-1,0)_L]_{55}$  \\
&&& $S^\alpha$         & $2\times[({\bf 2},{\bf 1},{\bf 1};{\bf 2},{\bf 1},{\bf 1})
(+1,0,0;+1,0,0)_L]_{95}$\\
    &&& $T^\alpha$       & $2\times [({\bf 1},{\bf 2},{\bf 1};{\bf 1},{\bf 1},{\bf 4})
(0,+1,0;0,0,+1)_L]_{95}$ \\
&&& $U^\alpha$ & $2\times [({\bf 1},{\bf 1},{\bf 4};{\bf 1},{\bf 2},{\bf 1})
(0,0,+1;0,+1,0)_L]_{95}$\\
 &&& ${\widetilde S}^\alpha$         & $2\times [({\bf 1},{\bf 2},{\bf 1};{\bf 1},{\bf 2},
{\bf 1})(0,-1,0;0,-1,0)_L]_{95}$  \\ 
     &&&${\widetilde T}^\alpha$      & $2\times [({\bf 2},{\bf 1},{\bf 1};{\bf 1},{\bf 1},
\overline{\bf 4})(-1,0,0;0,0,-1)_L]_{95}$\\ 
&&& ${\widetilde U}^\alpha$ & $2\times [({\bf 1},{\bf 1},{\overline {\bf 4}};{{\bf 2}},{\bf 1},{\bf 1})
(0,0,-1;-1,0,0)_L]_{95}$\\
\hline
\end{tabular}
\caption{The massless open string spectra of the ${\cal N}=1$ orientifolds of Type IIB on $T^6/
{\bf Z}_6$.
The $U(1)$ charges are
given in parentheses.}
\label{Z6} 
\end{table}

\begin{table}[t]
\begin{tabular}{|c|c|c|c|l|}
 Model & $b$ &Gauge Group  & Field & Charged Matter  
  \\
 \hline
${\bf Z}_6$ & 4 & $[U(2)\otimes U(2)]_{99}\otimes$  &  
 $\Phi_{1,2}$ &
 $2\times [({\bf 1},{\bf 1};{\bf 1},{\bf 1})(+2,0;0,0)_L]_{99}$
 \\
            &&  $[U(2)\otimes U(2)]_{55}$ 
 &  ${\widetilde \Phi}_{1,2}$ &
 $2\times [({\bf 1},{{\bf 1}};{\bf 1},{\bf 1})(0,-2;0,0)_L]_{99}$  \\
 &&& $R$ &$ [({\bf 2},{{\bf 2}};
     {\bf 1},{\bf 1})(+1,-1;0,0)_L]_{99}$  \\
 & &   &  
 $\Phi^\prime_{1,2}$ &
 $2\times [({\bf 1},{\bf 1};{\bf 1},{\bf 1})(0,0;+2,0)_L]_{55}$
 \\
            &&   
 &  ${\widetilde \Phi}^\prime_{1,2}$ &
 $2\times [({\bf 1},{\bf 1};{\bf 1},{{\bf 1}})(0,0;0,-2)_L]_{55}$  \\
 &&& $R^\prime$ &$  [(
    {\bf 1},{\bf 1};{\bf 2},{{\bf 2}})(0,0;+1,-1)_L]_{55}$  \\
&&& $S^\alpha$         & $4\times[({\bf 2},{\bf 1};{\bf 2},{\bf 1})
(+1,0;+1,0)_L]_{95}$\\
 &&& ${\widetilde S}^\alpha$         & $4\times[({\bf 1},{\bf 2};{\bf 1},{\bf 2})(0,-1;0,-1)_L]_{95}$  \\ 
\hline
 & 6 & $[U(2)]_{99}\otimes[U(2)]_{55}$  &  
  &
 none
 \\
\hline
${\bf Z}_2\otimes{\bf Z}_2\otimes {\bf Z}_3$ & 0 & $[U(6)\otimes Sp(4)]_{99}\otimes$  &  
 $\Phi_r$ &
 $3\times [({\bf 15},{\bf 1})(+2)_L]_{99}$
 \\
            &&  $\bigotimes_{s=1}^3 [U(6)\otimes Sp(4)]_{5_s 5_s}$ 
 &  $\chi_r$ &
 $3\times [({\overline {\bf 6}},{{\bf 4}})(-1)_L]_{99}$  \\
 
 &  &   &  
 $\Phi^s_r$ &
 $3\times [({\bf 15}_s,{\bf 1}_s)(+2_s)_L]_{99}$
 \\
            &&   
 &  $\chi^s_r$ &
 $3\times [({\overline {\bf 6}}_s,{{\bf 4}}_s)(-1_s)_L]_{5_s 5_s}$  \\
            &&   
 &  $P^s$ &
 $[({\bf 6},{\bf 1};{\bf 6}_s,{{\bf 1}}_s)(+1;+1_s)_L]_{95_s}$  \\
            &&   
 &  $Q^s$ &
 $[({\overline {\bf 6}},{\bf 1};{\bf 1}_s,{{\bf 4}}_s)(-1;0_s)_L]_{95_s}$  \\
            &&   
 &  $R^s$ &
 $[({\bf 1},{{\bf 4}};{\overline {\bf 6}}_s,{\bf 1}_s)(0;-1_s)_L]_{95_s}$  \\
 &&&  $P^{ss^\prime}$ &
 $[({\bf 6}_s,{\bf 1}_s;{\bf 6}_{s^\prime},{{\bf 1}}_{s^\prime}) (+1_s;+1_{s^\prime})_L]_{5_s
 5_{s^\prime}}$  \\
            &&   
 &  $Q^{ss^\prime}$ &
 $[({\overline {\bf 6}}_s,{\bf 1}_s;{\bf 1}_{s^\prime},{{\bf 4}}_{s^\prime})(-1_s;0_{s^\prime})_L]_{5_s 5_{s^\prime}}$  \\
            &&   
 &  $R^{ss^\prime}$ &
 $[({\bf 1}_s,{{\bf 4}}_s;{\overline {\bf 6}}_{s^\prime},{\bf 1}_{s^\prime})(0_s;-1_{s^\prime})_L]_{5_s5_{s^\prime}}$  \\
\hline
& 2 & $[U(2)\otimes Sp(4)]_{99}\otimes$  &  
 $\Phi_r$ &
 $3\times [({\bf 3},{\bf 1})(+2)_L]_{99}$
 \\
            &&  $\bigotimes_{s=1}^3 [U(2)\otimes Sp(4)]_{5_s 5_s}$ 
 &  $\chi_r$ &
 $3\times [({{\bf 2}},{{\bf 4}})(-1)_L]_{99}$  \\
 
 &  &   &  
 $\Phi^s_r$ &
 $3\times [({\bf 3}_s,{\bf 1}_s)(+2_s)_L]_{99}$
 \\
            &&   
 &  $\chi^s_r$ &
 $3\times [({{\bf 2}}_s,{{\bf 4}}_s)(-1_s)_L]_{5_s 5_s}$  \\
            &&   
 &  $P^{\alpha s}$ &
 $2\times [({\bf 2},{\bf 1};{\bf 2}_s,{{\bf 1}}_s)(+1;+1_s)_L]_{95_s}$  \\
            &&   
 &  $Q^{\alpha s}$ &
 $2\times [({{\bf 2}},{\bf 1};{\bf 1}_s,{{\bf 4}}_s)(-1;0_s)_L]_{95_s}$  \\
            &&   
 &  $R^{\alpha s}$ &
 $2\times [({\bf 1},{{\bf 4}};{{\bf 2}}_s,{\bf 1}_s)(0;-1_s)_L]_{95_s}$  \\
 &&&  $P^{\alpha ss^\prime}$ &
 $2\times [({\bf 2}_s,{\bf 1}_s;{\bf 2}_{s^\prime},{{\bf 1}}_{s^\prime}) (+1_s;+1_{s^\prime})_L]_{5_s
 5_{s^\prime}}$  \\
            &&   
 &  $Q^{\alpha ss^\prime}$ &
 $2\times [({{\bf 2}}_s,{\bf 1}_s;{\bf 1}_{s^\prime},{{\bf 4}}_{s^\prime})(-1_s;0_{s^\prime})_L]_{5_s 5_{s^\prime}}$  \\
            &&   
 &  $R^{\alpha ss^\prime}$ &
 $2\times [({\bf 1}_s,{{\bf 4}}_s;{{\bf 2}}_{s^\prime},{\bf 1}_{s^\prime})(0_s;-1_{s^\prime})_L]_{5_s5_{s^\prime}}$  \\
\hline
& 4 & $[U(2)]_{99}\otimes \bigotimes_{s=1}^3 [U(2)]_{5_s 5_s}$  &  
 $\Phi_r$ &
 $3\times [{\bf 1}(+2)_L]_{99}$
 \\
            &&   
 &   $\Phi^s_r$ &
 $3\times [{\bf 1}_s (+2_s)_L]_{99}$
 \\
            &&   
 &  $P^{\alpha s}$ &
 $4\times [({\bf 2};{\bf 2}_s)(+1;+1_s)_L]_{95_s}$  \\
 && &  $P^{\alpha ss^\prime}$ &
 $4\times [({\bf 2}_s;{\bf 2}_{s^\prime}) (+1_s;+1_{s^\prime})_L]_{5_s
 5_{s^\prime}}$  \\
 \hline
& 6 & $[Sp(2)]_{99}\otimes\bigotimes_{s=1}^3 [Sp(2)]_{5_s 5_s}$  &  
  &
 none
 \\
 \hline
\end{tabular}
\caption{The massless open string spectra of the ${\cal N}=1$ orientifolds of Type IIB on $T^6/
{\bf Z}_6$ and $T^6/{\bf Z}_2\otimes{\bf Z}_2\otimes {\bf Z}_3$.
The $U(1)$ charges are
given in parentheses.}
\label{Z6a} 
\end{table}

\begin{table}[t]
\begin{tabular}{|c|c|c|c|l|}
 Model &$b$&Gauge Group  & Field & Charged Matter  
  \\
 \hline
$T^6/{\cal G}$ & 0& $U(4)\otimes SO(8)$  &$ P$&
 $({\bf 4},{\bf 8}_v)(+1)_L$
  \\
  &&& $Q_{1,2}$ & $2\times ({\bf6},{\bf 1})(-2)_L$ \\
    &&& $\Phi$ & $({\overline{\bf 10}},{\bf 1})(-2)_L$\\  
\hline
& 6& $Sp(4)$ & &none\\
\hline
${\bf C}^3/{\cal G}$ && $U(3N-2\eta)\otimes U(N)^4\otimes $  &$ \Phi$&
 $({\bf R}_\eta,{\bf 1},{\bf 1},{\bf 1},{\bf 1},{\bf 1})(+2,0,0,0,0)_L$
  \\
  &&$G_\eta(N-2\eta)$& $Q$ & $({\bf A},{\bf 1},{\bf 1},{\bf 1},{\bf 1},{\bf 1})(+2,0,0,0,0)_L$ \\
    &&& $R$ & $({\bf S},{\bf 1},{\bf 1},{\bf 1},{\bf 1},{\bf 1})(+2,0,0,0,0)_L$\\
 &&& $\chi$ & $({\overline{\bf 3N-2\eta}},{\bf 1},{\bf 1},{\bf 1},{\bf 1},
 {\bf N-2\eta})(-1,0,0,0,0)_L$\\   
 &&& $P_1$ & $({\overline{\bf 3N-2\eta}},{\bf N},{\bf 1},{\bf 1},{\bf 1},{\bf 1})(-1,+1,0,0,0)_L$\\
     &&& ${\widetilde P}_1$ & $({\overline{\bf 3N-2\eta}},{\overline {\bf N}},{\bf 1},{\bf 1},{\bf 1},{\bf 1})(-1,-1,0,0,0)_L$\\  
 & && $P_{2,3,4},{\widetilde P}_{2,3,4} $ &plus cyclic permutations of the \\
 && && $U(N)\otimes U(N)\otimes U(N)\otimes U(N)$ irreps\\  
\hline
\end{tabular}
\caption{The massless open string spectra of the 
${\cal N}=1$ $\Omega$ orientifold of Type IIB on $T^6/{\cal G}$ discussed in subsection A of section IV, and the
${\cal N}=1$ $\Omega J$ orientifold of Type IIB on ${\bf C}^3/{\cal G}$ discussed in subsection B of section IV.  
The $U(1)$ charges are
given in parentheses. Here ${\bf A}$ and ${\bf S}$ stand for the two-index antisymmetric
and symmetric representations of the corresponding unitary group, respectively. Also,
${\bf R}_\eta={\bf A}$ for $\eta=-1$, ${\bf R}_\eta={\bf A}$ for $\eta=+1$, $G_\eta=SO$ for 
$\eta=-1$, and $G_\eta=Sp$ for $\eta=+1$.}
\label{nonabe} 
\end{table}

\end{document}